\documentclass[a4paper,12pt]{article}
\pdfoutput=1
\usepackage{epsfig}
\usepackage{amssymb,subfigure}
\usepackage{amsfonts}
\usepackage{amsmath}
\usepackage{euscript}
\usepackage{verbatim}
\usepackage{latexsym}
\usepackage{graphicx,color}
\usepackage{caption}
\usepackage{float}
\usepackage{slashed}
\usepackage{graphicx}
\graphicspath{ {images/} }
\usepackage{ytableau}
\usepackage{wrapfig}
	\usepackage[T1]{fontenc}
	\usepackage{tikz}
	\usetikzlibrary{decorations.pathmorphing}
\usepackage{tikz}
\usetikzlibrary{shapes.geometric, arrows,patterns,snakes}
\tikzstyle{ellip} = [ellipse, minimum width=3cm, minimum height=1cm,text centered, draw=black]

\newskip\humongous \humongous=0pt plus 1000pt minus 1000pt

\newif\ifdtup

\allowdisplaybreaks[1]

\catcode`\@=11

\@addtoreset{equation}{section}

\def\@normalsize{\@setsize\normalsize{15pt}\xiipt\@xiipt
\abovedisplayskip 14pt plus3pt minus3pt%
\belowdisplayskip \abovedisplayskip
\abovedisplayshortskip \z@ plus3pt%
\belowdisplayshortskip 7pt plus3.5pt minus0pt}

\def\small{\@setsize\small{13.6pt}\xipt\@xipt
\abovedisplayskip 13pt plus3pt minus3pt%
\belowdisplayskip \abovedisplayskip
\abovedisplayshortskip \z@ plus3pt%
\belowdisplayshortskip 7pt plus3.5pt minus0pt
\def\@listi{\parsep 4.5pt plus 2pt minus 1pt
     \itemsep \parsep
     \topsep 9pt plus 3pt minus 3pt}}

\relax

\catcode`@=12

\catcode`\@=11

\def\section{\@startsection{section}{1}{\z@}{3.5ex plus 1ex minus
   .2ex}{2.3ex plus .2ex}{\large\bf}}

\def\SymBoxes#1#2#3#4{\newdimen\un@t \un@t#3%
\raisebox{#1}{\rule{#2\un@t}{#4}\hskip-#2\un@t
\@tempdimb\un@t \advance\@tempdimb by-#4\@tempcntb#2\relax%
\@whilenum{\@tempcntb>0}\do{
\rule{#4}{\un@t}\hskip\@tempdimb \advance\@tempcntb by\m@ne}%
\hskip-#2\un@t \rule[\un@t]{#2\un@t}{#4}%
\rule[\un@t]{#4}{#4}\hskip-#4
\rule{#4}{\un@t}}\hskip-#4}                

\begin{document}

\newcommand{\beq}{\begin{equation}}
\newcommand{\eeq}{\end{equation}}
\newcommand{\bea}{\begin{eqnarray}}
\newcommand{\eea}{\end{eqnarray}}
\newcommand{\beas}{\begin{eqnarray*}}
\newcommand{\eeas}{\end{eqnarray*}}
\newcommand{\defi}{\stackrel{\rm def}{=}}
\newcommand{\non}{\nonumber}
\newcommand{\bquo}{\begin{quote}}
\newcommand{\enqu}{\end{quote}}
\renewcommand{\(}{\begin{equation}}
\renewcommand{\)}{\end{equation}}
\def \eqn#1#2{\begin{equation}#2\label{#1}\end{equation}}
\def\IZ{{\mathbb Z}}
\def\IR{{\mathbb R}}
\def\IC{{\mathbb C}}
\def\IQ{{\mathbb Q}}
\def\de{\partial}
\def\Tr{ \hbox{\rm Tr}}
\def\H{ \hbox{\rm H}}
\def\HE{ \hbox{$\rm H^{even}$}}
\def\HO{ \hbox{$\rm H^{odd}$}}
\def\K{ \hbox{\rm K}}
\def\Im{ \hbox{\rm Im}}
\def\Ker{ \hbox{\rm Ker}}
\def\const{\hbox {\rm const.}}
\def\o{\over}
\def\im{\hbox{\rm Im}}
\def\re{\hbox{\rm Re}}
\def\bra{\langle}\def\ket{\rangle}
\def\Arg{\hbox {\rm Arg}}
\def\Re{\hbox {\rm Re}}
\def\Im{\hbox {\rm Im}}
\def\exo{\hbox {\rm exp}}
\def\diag{\hbox{\rm diag}}
\def\longvert{{\rule[-2mm]{0.1mm}{7mm}}\,}
\def\a{\alpha}
\def\dag{{}^{\dagger}}
\def\tq{{\widetilde q}}
\def\p{{}^{\prime}}
\def\W{W}
\def\N{{\cal N}}
\def\hsp{,\hspace{.7cm}}

\def\br{\nonumber\\}
\def\IZ{{\mathbb Z}}
\def\IR{{\mathbb R}}
\def\IC{{\mathbb C}}
\def\IQ{{\mathbb Q}}
\def\IP{{\mathbb P}}
\def \eqn#1#2{\begin{equation}#2\label{#1}\end{equation}}

\newcommand{\sgm}[1]{\sigma_{#1}}
\newcommand{\idd}{\mathbf{1}}

\newcommand{\C}{\ensuremath{\mathbb C}}
\newcommand{\Z}{\ensuremath{\mathbb Z}}
\newcommand{\R}{\ensuremath{\mathbb R}}
\newcommand{\rp}{\ensuremath{\mathbb {RP}}}
\newcommand{\cp}{\ensuremath{\mathbb {CP}}}
\newcommand{\vac}{\ensuremath{|0\rangle}}
\newcommand{\vact}{\ensuremath{|00\rangle}                    }
\newcommand{\oc}{\ensuremath{\overline{c}}}
\begin{titlepage}
\begin{flushright}
CHEP XXXXX
\end{flushright}
\bigskip
\def\thefootnote{\fnsymbol{footnote}}

\begin{center}
{\Large
{\bf Circuit Complexity in Fermionic Field Theory
}
}
\end{center}

\bigskip
\begin{center}
{\large  Rifath KHAN$^a$\footnote{\texttt{rifathkhantheo@gmail.com}}, Chethan KRISHNAN$^a$\footnote{\texttt{chethan.krishnan@gmail.com}}, Sanchita SHARMA$^a$\footnote{\texttt{sanchitas39@gmail.com}} \vspace{0.15in} \\ }
\vspace{0.1in}

\end{center}

\renewcommand{\thefootnote}{\arabic{footnote}}

\begin{center}
$^a$ {Centre for High Energy Physics,\\
Indian Institute of Science, Bangalore 560012, India}

\end{center}

\noindent
\begin{center} {\bf Abstract} \end{center}
We define and calculate versions of complexity for free fermionic quantum field theories in 1+1 and 3+1 dimensions, adopting Nielsen's geodesic perspective in the space of circuits. We do this both by discretizing and identifying appropriate classes of Bogoliubov-Valatin transformations, and also directly in the continuum by defining squeezing operators and their generalizations. As a closely related problem, we consider cMERA tensor networks for fermions: viewing them as paths in circuit space, we compute their path lengths. Certain ambiguities that arise in some of these results because of cut-off dependence are discussed. 




\vspace{1.6 cm}
\vfill

\end{titlepage}

\setcounter{page}{2}
\tableofcontents

\setcounter{footnote}{0}


\section{Introduction}

The formation of black holes via gravitational collapse in anti-de Sitter space is expected to be dual to thermalization in the dual conformal field theory. This leads one to think of a thermal CFT State as a gravitational configuration that can be approximated by an eternal black hole. But if this is so, one needs to have a CFT explanation for the fact that the Einstein-Rosen bridge (or wormhole) that shows up inside the horizon of an eternal black hole is a time-dependent configuration, and that its "size" increases towards the future. It has recently been proposed by Susskind and others \cite{coredump} that the quantity one should compare against the size of the Einstein-Rosen wormhole is the "complexity" of the CFT State: the idea being that the complexity of a state can increase even after it has thermalized in some appropriate sense.


The trouble however is that the definition of complexity is often very murky. Loosely it captures the number of bits that are required to describe a state. For example, when coffee and milk are separate, the complexity is comparatively minimal. But once they start mixing, one expects that the complexity increases for a while, but after a while it starts to go down: once the mixing is complete, the complexity is again minimal. So while entropy and the thermodynamic arrow of time increase forever, the qualitative expectation is that complexity first increases and then decreases. That said, a uniquely compelling quantitative definition of complexity that manifests this behaviour is not known, to the best of our knowledge. 

One way to make things a bit more concrete is to consider "circuit complexity", where one assigns complexity to quantum states as follows. First one picks a reference state and a set of unitary operators that are called gates.  Then the complexity of any particular state is captured by the minimal number of gates that one must act with on the reference state in order to get to that particular state. This is believed to be a fairly reasonable definition of complexity, even though clearly it involves some arbitrary choices: what is a good reference state? What are good choices for the gate unitaries? ... \footnote{Note also that this definition is usually implemented in discrete qubit systems and often with discrete time evolution \cite{susskind-neg}.}. For our purposes in this paper, it is worth noting that this minimization of the number of gates can be re-interpreted \`a la Nielsen as a geodesic length minimization in the space of circuits \cite{Nielsen}. 

To make a sensible holographic definition of complexity from the CFT, one needs to generalize these constructions in three substantive ways. Firstly, we need a definition that generalizes these issues to the setting of quantum field theory rather than in the setting of qubits and 0+1 dimensional quantum mechanics, as is usual. Secondly, since at least in known concrete examples of holography the boundary theory has a gauge invariance, it seems necessary that we will need to define complexity in the context of  gauge theories, in particular non-Abelian gauge theories. Thirdly, perhaps after some field theory constructions have been undertaken for the free theory, one would need to come up with a way to make such a definition viable, for strongly coupled theories. After all, weakly coupled gravitational physics is captured by strongly coupled field theory. 

Recently, two related but different attempts at the first of these problems was made in the context of scalar field theories in \cite{Myers, Chapman}. In this paper, we will consider both these approaches and adapt and generalize these approaches to include various classes of fermionic field theories. One of our motivations for this undertaking is the recent emergence of a class of strongly coupled fermionic gauged quantum mechanical theories called SYK-like tensor models, as well as their generalizations to higher dimensions which are bonafide field theories. Of course, this paper does not deal with neither gauge theories nor strongly coupled systems: it should be viewed as a preliminary exploratory attempt. We hope that our explicit calculations will be useful in shedding some light on the questions at the intersection of holography and complexity, and perhaps in sharpening them. 

The structure of the calculation in \cite{Myers} was to discretize spacetime and then work effectively with quantum mechanical wave functions for the discretized oscillators. By working with a class of Gaussian wave functions that interpolated between the decoupled reference state and the true ground state (chosen as the target state) they were able to define complexity for the state by minimizing the Nielsen-like path length in the space of unitaries that did such an interpolation. Since we are working with fermions, an analogue of the Gaussian wave function will involve Grassmann variables and will be too awkward for many purposes. But we will see that it is easy enough to work directly with (finite dimensional) fermionic Hilbert spaces that contain the reference and target states, and that they lead to natural notions of circuit length minimization. The slight subtlety due to fermion doubling in the lattice will not affect the main points we make. As a result we find an expression for the complexity, which is controlled by the formula in \eqref{main D}\footnote{After this result was obtained by us, a paper \cite{Ross} appeared on the arXiv which considers discretized complexity for Dirac fermions. Even though we work with Majorana spinors and they seem to be using Dirac, it is reasonable for the results to be similar: we present concrete arguments why this is so (but in the continuum case), in section 3. One of our results for the discrete case \eqref{main D} can be re-written as $ \cos^{-1} \sqrt{\frac{1}{2}+\frac{\omega}{\sqrt{4 \omega^{2}+\Omega^{2}}}} $, which we suspect is the same as eqn. (73) in \cite{Ross}. See also the related earlier work \cite{Preskill}.}, and  which shows up in various guises in the various cases we consider. This fermionic form is distinct from the form of complexity in \cite{Myers}, where the analogous expression is a $\log$ instead of our $\arctan$. 

On the continuum side, we will find that a construction entirely parallel to that of \cite{Chapman} for scalars is possible for Dirac fermions in 1+1 dimensions. The strategy of using squeezing operators and their generalizations as entangler operators turns out to be a viable strategy for fermions as well. Our analysis of the 1+1 dimensional continuum Dirac case works entirely parallel to the results for the scalar in \cite{Chapman}. Instead of the metric on the hyperbolic plane, we discover the metric on a $\C \IP^1$ sphere, and instead of an $SU(1,1)$ isometry, we find an $SU(2)$. But we further generalize our approach to Majorana theory in 1+1 dimensions, as well as to massive and massless Dirac theories in 3+1 dimensions. In all these cases we are able to identify convenient squeezing operators that take us to natural target states that approximate the ground state. 

In the final section before the conclusion, we discuss the cMERA tensor network for fermions, as an alternate path in the space of unitaries. We first do this for 1+1 dimensions and then move on to 3+1 dimensions (where we introduce a new cMERA-like path), both in the massive and massless cases. In both cases we find that as the cut-off tends to infinity, the state tends to the ground state. But since {\em when the cut off is finite} the target state for the cMERA is not quite the target state of the previous paragraph (even though they both tend to the ground state as $\Lambda \rightarrow \infty$), it is not possible to meaningfully compare the lengths of the paths at finite cut-off by looking at their leading divergences. This leads us to a discussion of the meaning of a cut-off dependent quantity like the complexity, and to speculations on the possibility that sub-leading terms in complexity could be of physical interest, in analogy with (holographic) entanglement entropy. In particular, these UV divergences could be reinterpreted as AdS IR divergences in a holographic context. But it must be born in mind here that we are dealing with free quantum field theories in our work.

In a concluding section we make various speculative comments. In particular, we discuss the possibility of doing similar calculations for gauge theories and perturbative string theory. Especially since the worldsheet theory is free after gauge fixing, the latter is tractable and is currently under investigation.  We also briefly allude to some issues which were suppressed in the main text, the choice of metric and penalty factors in Nielsen's geometric definition among others. Various appendices contain some review material as well as technical details.

\section{Lattice of Fermionic Oscillators}

Lets us start with the Dirac Lagrangian in $d+1$ spacetime dimensions: 
\begin{equation} 
L =  \overline{ \Psi}( i \gamma . \partial - m ) \Psi  \end{equation} 
 The conjugate momentum is 
\begin{equation} 
\Pi = \frac{ \partial L}{ \partial ( \partial_{0} \Psi)} = i \Psi^{\dagger} 
\end{equation} 
and the Dirac hamiltonian is 
\begin{equation}  
H = \int d^{d}x [ -i \overline{ \Psi} \gamma^{i} \partial_{i} \Psi +m \overline{ \Psi} \Psi ] 
\end{equation}
We regulate it by placing it on a lattice with lattice spacing $ \delta $ and the Hamiltonian becomes 
\bea 
H = \underset{\vec{n}}{\sum} \delta^{d} \left[ -i \overline{ \Psi} ( \vec{n}) \sum_i\gamma^{i} \Big( \frac{ \Psi ( \vec{n}+\hat{x}_i) - \Psi ( \vec{n})}{ \delta}\Big) + m \overline{\Psi} ( \vec{n}) \Psi( \vec{n}) \right] 
\eea
the $x_i$ here are "unit''\footnote{The quotes around the "unit'' emphasise that the length of the vector is the lattice spacing.} vectors along the axes of the ($d$-dimensional) lattice. The summation over $i$ is only over the spatial directions, and therefore range over $d$ values. By introducing $ \omega \equiv \delta^{d} m $ and $ \Omega \equiv \delta^{d-1} $ we can bring it to the form
\bea
\underset{\vec{n}}{\sum} \left[ -i \Omega \ \overline{ \Psi} ( \vec{n}) \sum_i\gamma^{i} (\Psi ( \vec{n}+\hat{x}_i) - \Psi ( \vec{n})) + \omega \overline{\Psi} ( \vec{n}) \Psi( \vec{n}) \right] .\label{discreteO}
\eea
Note that this is a system of coupled fermionic oscillators, but since the coupling is quadratic, we can solve the system completely by identifying the normal modes. We will see various versions of this general idea throughout this paper.

\subsection{A Toy Model}

The above form of the Hamiltonian suggests that  we should study a set of coupled oscillators of the form
\begin{equation} 
H = \omega ( \overline{ \Psi}_{1} \Psi_{1} +  \overline{ \Psi}_{2} \Psi_{2}) + i \Omega \ \overline{ \Psi}_1 \rho^{1} ( \Psi_1 - \Psi_2). \label{Minimal}
\end{equation}
The basic idea is this: we wish to consider the simplest oscillator system which has a coupling analogous to the discretisation \eqref{discreteO} of the field theory. A candidate is the system with two lattice points parallel to the work for scalars considered in \cite{Myers}. But if we work with $d+1$-dimensional Dirac fermions, the coupling term will force us to consider the $d$ spatial directions. We wish to avoid this technical complication, and therefore in writing the above, we have restricted ourselves to a $1+1$-dimensional theory. After we gain some intuition by working with this system, we will consider more general cases. In particular, we will see that the above Hamiltonian \eqref{Minimal}, when interpreted as a theory for Majorana fermions, is essentially a precise parallel to the bosonic two oscillator case discussed in \cite{Myers}. We have adopted the notation $\rho^a$ for gamma matrices in two dimensions in this subsection. 

In two dimensions, gamma matrices allow a purely imaginary representation in which the spinors can be taken as Majorana. The gamma matrices take the form
\bea
\rho^0=\left(
\begin{array}{cc}
 0 & -i  \\
 i & 0  \\
\end{array}
\right), \ \ 
\rho^1=\left(
\begin{array}{cc}
 0 & i  \\
 i & 0  \\
\end{array}
\right)
\eea
The general Majorana spinor in two dimensions $\Psi_i$ (here $i$ is part of the name of the spinor and not a spinor index) can be written as
\bea
\Psi_i=\left(
\begin{array}{c}
 \psi_i^-   \\
 \psi_i^+  \\
\end{array}
\right), 
\eea
where the $\psi_i^{\pm}$ are real Grassmann variables. This means that our minimal Hamiltonian \eqref{Minimal} is made of four real Majorana fermions and takes the form
\bea
H=2i \omega (\psi_1^+\psi_1^-+\psi_2^+\psi_2^-)+i \Omega(\psi_1^+\psi_2^+-\psi_1^-\psi_2^-).
\eea
Note that the Hamiltonian is real (Hermitian) because the fermions are Majorana and anti-commute. Now we will re-write this Hamiltonian in a form that is useful for future generalizations\footnote{Note that because the fermions are Majorana, we can use Hermitian conjugation and transposition interchangeably.}.

We will define complex Grassmann variables $ b_{i} $ and $ b^{\dagger}_{i} $
\bea
b_{i} = \frac{\psi^{-}_{i} - i \psi^{+}_{i}}{\sqrt{2}} \quad 
b^{\dagger}_{i} = \frac{\psi^{-}_{i} + i \psi^{+}_{i}}{\sqrt{2}}
\eea
and quantize by imposing the anti commutation relations $ \{b_{i}, b^{\dagger}_{j} \} = \delta_{ij} $, see Appendix A for our conventions on fermionic oscillators. The Hamiltonian takes the form 
\bea
H = \omega [b^{\dagger}_{1} , b_{1}] + \omega [b^{\dagger}_{2} , b_{2}] - i \Omega (b^{\dagger}_{1} b^{\dagger}_{2} + b_{1}b_{2})
\eea
To bring this to the diagonalised normal mode form, we use the standard technology of Bogoliubov-Valatin (BV) transformations (see Appendix B for a self-contained review). The result is
\bea
H = \lambda [\tilde{b}^{\dagger}_{1} , \tilde{b}_{1}] +  \lambda [\tilde{b}^{\dagger}_{2} , \tilde{b}_{2}]
\eea
where
\bea
\lambda = \frac{1}{2} \sqrt{4 \omega^{2} + \Omega^{2}}.
\eea 

\subsection{Complexity of the Toy Model Ground State}

A B-V transformation matrix that does the job of diagonalising the Hamiltonian in the previous section is 
\bea
T = \left[ \begin{array} {cccc}
0 & \eta & 
i\rho & 0 \\ \eta & 0 & 0 &  - i \rho  \\
 -i \rho & 0& 0 & \eta \\ 0 &  i\rho  & \eta &0
\end{array} \right ]
\eea
where
\bea
\rho=  \sqrt{\frac{1}{2}-\frac{\omega}{\sqrt{4 \omega^{2}+\Omega^{2}}}},
\qquad 
\eta= \sqrt{\frac{1}{2}+\frac{\omega}{\sqrt{4 \omega^{2}+\Omega^{2}}}}
\eea
This form suggests that we define 
\bea
r\equiv \frac{1}{2} \tan^{-1}\left( \frac{\Omega}{2\omega}\right), \label{main D}
\eea
so that 
\bea
\rho = \sin r, \ {\rm and} \ \  \eta = \cos r.
\eea

To define complexity, we need to define a reference state from which we reach the target states via appropriate unitary transformations. A natural reference state is to choose $|00 \rangle$, defined via
\bea
b_{i} |00 \rangle = 0, \quad {\rm for} \quad i = 1,2,
\eea
and the target state which is the ground state of our theory is defined by
\bea
\tilde{b_{i}} | \tilde{0} \tilde{0} \rangle = 0, \quad {\rm for} \quad i = 1,2.
\eea
From the B-V matrix, it follows that
\bea | \tilde{0} \tilde{0} \rangle = \cos r \ |00 \rangle + i \sin r\ | 11 \rangle
\eea 
To go from the reference state to the target state by a unitary transformation, we first write the states in matrix form with the basis $|ij \rangle$. The unitary transformation is a $4 \times 4$ unitary matrix:
\bea
\left[ \begin{array} {c}
\cos r \\ 0 \\ 0 \\ i \sin r \end{array} \right ]
= U^{(4\times 4)}
\left [\begin{array} {c}
1 \\ 0 \\ 0 \\ 0 \end{array} \right ]
\eea
Notice that the middle two components of the state vector is zero for both reference and target state. If these two components are changed along the path then we need to bring them back down to zero, when we get to the target state. So it stands to reason that (for appropriately defined Euclidean notions of distance in the space of unitaries) this would only increase the length, so we restrict ourselves to the path which only changes the first and fourth components. These transformations clearly fall inside a $U(2)$ group.
\bea
\left[ \begin{array} {c}
\cos r \\ i \sin r\end{array} \right ]
= U^{(2\times 2)}
\left [\begin{array} {c}
1\\ 0 \end{array} \right ]
\eea
Now $U^{(2\times 2)}$ is a $2 \times 2$ unitary matrix. After extracting a $U(1)$ phase $e^{i y}$ and writing the remaining $SU(2)$ as an $S^3$, we can parametrise a general $U(2)$ matrix as
\bea
U=\left[ \begin{array} {cc}
e^{iy} \cos{\rho} \cos{\tau} + i e^{iy} \sin{\theta}\sin{\rho} & e^{iy} \cos{\theta} \sin{\rho} + i e^{iy} \cos{\rho} \sin{\tau} \\
ie^{iy} \cos{\rho} \sin{\tau} - e^{iy} \cos{\theta} \sin{\rho} & e^{iy} \cos{\rho} \cos{\tau} -i e^{iy}  \sin{\theta}\sin{\rho}
\end{array} \right ]
\eea
The ranges of the various coordinates are\footnote{We will work with $y \in [-\pi, \pi)$ when we want to go across the $y=0$ point without changing charts.}
\bea
y, \theta, \tau \in [0, 2 \pi), \ \ {\rm and} \ \ \rho \in [0, \pi/2].
\eea

For finite value of the parameter along the path, the state will be
\bea
| \Psi (\sigma) \rangle = U(\sigma)| R \rangle 
\eea
where  $ | R \rangle $ is the reference state. This means that in the explicit matrix above, we treat $\rho, \tau, \theta, y$ as functions of $\sigma$. We can view this unitary as a path-ordered exponential
\bea
U(\sigma) =  \overleftarrow{P} e^{ \int^{\sigma}_{0}{Y^{I}(s) O_{I}}} \label{pathordered}
\eea
if we write\footnote{This equation is the analogue of the time-dependent Schrodinger equation, written in a form that is usually written when one solves it via time-ordered exponentials. The left hand side is the analogue of the Hamiltonian. So the solution, \eqref{pathordered}, can be directly exported here by analogy. To write \eqref{Schrolike}, we merely note that since the right had side is made from a general $U(2)$ matrix, the left hand side must be writable in terms of the generators of $U(2)$ whose coefficients we call $Y^I$.}
\bea
Y^{I}(s)O_{I} = (\partial_{s} U(s)) U^{-1}(s) \label{Schrolike}
\eea
Here the  $ O_{I} $ are the generators of $U(2)$, and they can be taken as
\bea
\begin{split}
O_{0} = \left[ \begin{array} {cc}
i & 0 \\ o & i \end{array} \right], &\qquad &
O_{1} = \left[ \begin{array} {cc}
0 & i \\ i & 0 \end{array} \right], \\
O_{2} =\left [\begin{array} {cc} 
0 & 1 \\ -1 & 0 \end{array} \right], &\qquad &
O_{3} = \left[ \begin{array} {cc} i & 0 \\ 0 & -i \end{array} \right]. 
\end{split}
\eea
Using
\bea
{\rm Tr} (O_a \ O_b) =-\frac{1}{2} \delta_{ab},
\eea
we can extract the velocities via
\bea
Y^{I}(s) = - \frac{1}{2} \ {\rm Tr} \ [(\partial_{s} U(s))(U^{-1}(s) O_{I})]
\eea
Using these, we define the length of the path (using a Euclidean Metric) as
\bea
D[u] = \int^{1}_{0} ds \sqrt{\delta_{IJ}Y^{I}(s)Y^{J}(s)}
\eea
The path taken by the states will be a curve in $y, \rho, \tau, \theta $ coordinates. The $Y^I$'s will be linear in derivatives of them. 
Explicitly calculating this metric, we find the natural metric on $U(1) \times SU(2)$: 
\bea
ds^{2} = \sin^{2} (\rho) d \theta^{2} + d \rho^{2} + \cos^{2} (\rho) d \tau^{2} + dy^{2}
\eea
Extremizing this length gives us the geodesic equations
\bea
y'' = 0 \\ \nonumber
\rho''-\sin{\rho}\ \cos{\rho} \ \theta'^{2}  + \sin{\rho}\ \cos{\rho} \ \tau'^{2} = 0 \\ \nonumber
\tau'' - 2\ \tan{\rho} \ \rho'\ \tau' = 0 \\ \nonumber
\theta'' + 2 \ \cot{ \rho} \ \theta'\ \rho' = 0
\eea
where the derivatives are with respect to the curve parameter. 
Looking at the metric it is evident that $ \frac{\partial}{\partial y} , \frac{\partial}{\partial \theta} , \frac{\partial}{\partial \tau} $ are the killing vectors. Using these one sees that 
\bea
y'= {\rm const.},  \ \ 
\cos^2 \rho \ \tau'= {\rm const.}, \ \
\sin^2 \rho \ \theta'={\rm const.},
\eea
are the constants of motion. It is easy to check that this same result can be obtained by directly integrating the relevant equations of motion.

Now we can use the boundary conditions to solve the system. Demanding
\bea
U(s =0) = I
\eea
immediately yields 
\bea
(y(s=0), \rho(s=0) , \tau(s=0) , \theta(s=0)) = (0,0,0,\theta_{0}).
\eea
Furthermore, we know that
\bea
|\psi_{T} \rangle = U(s =1) |R \rangle.
\eea
Here $|\psi_{T} \rangle$ is the target state (and our target state is the ground state). A $2 \times 2$ unitary matrix that takes the reference state to the target state is of the general form
\bea
U(s =1) =\left[ \begin{array} {cc}
\cos r & i \sin r \ e^{-i\phi} \\
i \sin r & \cos r \ e^{-i\phi}
\end{array} \right ]=e^{-i\phi/2}\left[ \begin{array} {cc}
\cos r \ e^{i\phi/2} & i \sin r \ e^{-i\phi/2} \\
i \sin r\ e^{i\phi/2} & \cos r \ e^{-i\phi/2}
\end{array} \right ]
\eea
for some arbitrary $\phi \in [-\pi, \pi)$. This translates to the end point boundary condition in terms of coordinates
\bea (y(s=1), \rho(s=1), \tau(s=1), \theta(s=1) ) = (y_{1} , \rho_{1} , \tau_{1}, \theta_{1}) \eea
where
\bea
y_{1} = -\phi/2
,\qquad
\tau_{1} = r
,\qquad
\rho_{1} = \phi/2
,\qquad \theta_1=\pi/2 -r.  
\eea
Because the metric is Euclidean, one can convince oneself that the distance with the above initial and final boundary conditions is globally minimised for $\phi=0$\footnote{Note that when this happens, $\theta$ is no longer determinate, including at the boundaries.}, and the corresponding geodesic is given by
\bea
y(s) = 0
,\qquad
\rho(s) =  0, \hspace{0.25in} \nonumber\\ 
\tau(s) = \tau_{1} s , \qquad
\theta(s) = {\rm unfixed} 
\eea
The length of this minimum path can be directly calculated and the result is
\bea
D[U] = \int^{1}_{0} ds \ \tau_{1} = r = 
 \frac{1}{2} \tan^{-1}\left( \frac{\Omega}{2\omega}\right). 
\eea
This minimum length is the complexity of the target state.

In the analogous discussion in \cite{Myers}, the generators (the $M^I$ in their notation) used were non-standard and that resulted in a more complicated form of the metric and resulting geodesic equations. To solve the geodesic equations, more Killing vectors were identified. Here on the other hand, we explicitly see the $S^3 \times \IR$  form of the metric, and we only needed to use the obvious Killing vectors to find the explicit solution.


\subsection{Squeezing Operators as Gates}

In this section, we will present an alternate approach for discussing complexity, which has a natural role in the continuum case. We will define an entangling operator $K$, which is also known as squeezing operator in some contexts:
\bea
K = b^{\dagger}_{2} b^{\dagger}_{1} - b_{2}b_{1}= \tilde{b}^{\dagger}_{1}\tilde{b}^{\dagger}_{2} - \tilde{b_{1}} \tilde{b_{2}}
\eea
The first equality can be viewed as the definition of the operator, the second equality is the result of a calculation, where we have used the definition of the tilde'd operators via the B-V transformation from the previous subsection. 

It is also useful to define a unitary
\bea
U=e^{-iKr}
\eea
where $r\in \IR$. The target state $| \tilde{0} \tilde{0} \rangle$ can be reached from the reference state $|00 \rangle $ via
\bea
| \tilde{0} \tilde{0} \rangle =  U |00 \rangle
\eea
for some appropriately chosen $r$. 
It is easy to see this via a similarity transformation of $\tilde{b_{i}}$ using $U$ that gives $b_{i}$: 
\bea
U^{\dagger} \tilde{b_{1}} U &=& e^{iKr} \tilde{b_{1}}e^{-ikr} \\ \nonumber
&=& \tilde{b_{1}} - ir \tilde{b}^{\dagger}_{2} - \frac{r^{2}}{2}\tilde{b_{1}} +\frac{ir^{3}}{3!} \tilde{b}^{\dagger}_{2} + \frac{1}{4!} r^{4}\tilde{b}_{1}... \\ \nonumber
&=& \tilde{b_{1}}\cos{r} - i\tilde{b}^{\dagger}_{2} \sin{r}
\eea
Setting $U^{\dagger} \tilde{b_{1}}U = b_{2}$, we see that the transformation $U$ does indeed take the reference state to the target state that we have worked with in the previous sections if we take the value of $r$ to be what we found before.

Now let's consider an arbitrary path generated by the squeezing operator which takes the reference state to some more general target state
\bea
|\Psi(\sigma)\rangle=U(\sigma)|00\rangle=e^{-i K Y(\sigma)}|00\rangle
\eea
such that 
\bea
Y(0)=0   \implies |\Psi(0)\rangle=|00\rangle \\ \nonumber
Y(1)=r    \implies |\Psi(1)\rangle=|\tilde{0} \tilde{0}\rangle
\eea
We evaluate the length of this path using Fubini-Study metric and minimising it to get the complexity. In other words, the allowed circuits we are considering are the ones generated by $K$.

The Fubini-Study metric is
\bea
ds_{FS} (\sigma) = d \sigma \sqrt{|\partial_{\sigma}|\Psi(\sigma)\rangle|^{2} - |\langle \Psi(\sigma)|\partial_{\sigma}|\Psi(\sigma)|^{2}}
\eea
The circuit length of a path traced by intermediate states $ | \Psi (\sigma) \rangle = U( \sigma) | 00 \rangle $ is
\bea
\ell(| \Psi(\sigma) \rangle) = \int_{\sigma_{i}}^{\sigma_{f}} ds_{FS}(\sigma)
\eea
The complexity is the length of minimal path.
\bea
C= \underset{ \{ Y(s) \} } { { \rm min}} \ \ell(ds_{FS})=
\underset{ \{ Y(\sigma) \} } {{ \rm min}} \int_{0}^{1} d\sigma \sqrt{(\partial_{\sigma} Y (\sigma))^{2}} 
\eea 
This gives the geodesic to be a straight line path. Under a simple affine parametrisation $\sigma $ the geodesic is
\bea
Y(\sigma) =\sigma r
\eea 
The complexity for the target state is the length of this straight line path
\bea
C = \int_{0}^{1} d\sigma \sqrt{(r)^{2}}=r. 
\eea
This matches with the complexity derived in the previous section.
The idea here is that the unitary transformation was generated by one generator $K$, the squeezing operator. We interpret the squeezing operator as creating entanglement between the two oscillators, and we use this as another approach to the definition of complexity. In the present case, we see that the two approaches match.

\subsection{1+1 Dimensional Majorana on a Lattice}

So far we have considered just a pair of fermionic oscillators, albeit with a coupling that was motivated by our eventual interest in field theory. Now we will consider discretized versions of the field theory, and consider full lattices, with periodic and anti-periodic boundary conditions. We will work with Majorana fermions for concreteness. 


The Hamiltonian of 1+1 d Majorana theory in terms of the complexified variables is (see section 3.5 for a derivation):
\bea
H = \int dx (-i \Psi \partial_{1} \Psi - i \Psi^{\dagger} \partial_{1} \Psi^{\dagger} + m[\Psi^{\dagger} , \Psi])
\eea
By placing it on a circular\footnote{We use the word circular to refer to both periodic and anti-periodic boundary conditions simultaneously. The temptation to attribute any other meaning to this adjective should be resisted.} lattice with lattice spacing $\delta$, the Hamiltonian for $N$ oscillators is
\bea
H = \sum^{N-1}_{n =0} \delta \Big( -i \Psi_{n}\frac{(\Psi_{n+1} - \Psi_{n})}{\delta} - i \Psi^{\dagger}_{n}\frac{(\Psi^{\dagger}_{n+1} - \Psi^{\dagger}_{n})}{\delta} + m[\Psi^{\dagger}_{n} , \Psi_{n}] \Big)
\eea
We can quantize by imposing the anti-commutation relations
\bea
\{\Psi_{n}, \Psi^{\dagger}_{m} \} = \delta_{nm} \qquad \{\Psi_{n}, \Psi_{m} \} =0 = \{\Psi^{\dagger}_{n}, \Psi^{\dagger}_{m} \}
\eea
The Hamiltonian will be written as
\bea
H = \sum^{N-1}_{n =0}  (\omega [\Psi^{\dagger}_{n} , \Psi_{n}] -i (\Psi_{n} \Psi_{n+1} +\Psi^{\dagger}_{n}\Psi^{\dagger}_{n+1}))
\eea
where $ \omega = m \delta$.
We will look at the lattice with periodic (Ramond) and anti-periodic (Neveu-Schwarz) boundary conditions.

\subsubsection{Ramond Boundary Condition}

Ramond boundary condition is imposed by
\bea
\Psi_{n+N} = \Psi_{n}
\eea
The Discrete fourier transform for this boundary condition is
\bea
\Psi_{n} = \frac{1}{\sqrt{N}} \sum^{N-1}_{k=0} e^{\frac{2 \pi i k n }{N}} \Psi_{k}
\eea
and the inverse discrete Fourier transform is
\bea
\Psi_{k} = \frac{1}{\sqrt{N}} \sum^{N-1}_{k=0} e^{-\frac{2 \pi i k n }{N}} \Psi_{n}.
\eea
From the sum rule of $n$-th roots of unity
\bea
\delta_{k,-k'} = \frac{1}{N} \sum^{N-1}_{n=0} e^{\frac{2 \pi i}{N}n(k+ k')}
\eea
one can show that the anti-commutation relations translate to
\bea
\{\Psi_{k}, \Psi^{\dagger}_{k^{\prime}} \} = \delta_{k k^{\prime}} \qquad \{\Psi_{k}, \Psi_{k^{\prime}} \} =0 = \{\Psi^{\dagger}_{k}, \Psi^{\dagger}_{k^{\prime}} \}
\eea
One can check that the Fourier transformed variables also satisfy a Ramond-like boundary condition in Fourier space:
\bea
\Psi_{k+N} = \Psi_{k} \label{fperiod}
\eea
Using all of these, the Hamiltonian in terms of the Fourier transformed variables can be written as
\bea
H = \sum^{N-1}_{k =0}  \big( \omega [\Psi^{\dagger}_{k} , \Psi_{k}] -i (\Psi_{k} \Psi_{-k}e^{\frac{-2 \pi ik}{N}} +\Psi^{\dagger}_{k}\Psi^{\dagger}_{-k} e^{\frac{2 \pi ik}{N}})  \big).
\eea
Since the range of $k$ is periodic, we can use $\Psi_{-k} = \Psi_{N-k}$ to bring this to a more convenient form:
\bea
H = \sum^{N-1}_{k =0}  \big( \omega [\Psi^{\dagger}_{k} , \Psi_{k}] -i (\Psi_{k} \Psi_{N-k}e^{\frac{-2 \pi ik}{N}} +\Psi^{\dagger}_{k}\Psi^{\dagger}_{N-k} e^{\frac{2 \pi ik}{N}}) \big)
\eea
This form can be directly translated to the normal modes by an adaptation of our earlier B-V construction. We can see that the oscillators at $k$ and $N-k$ are getting mixed, independently from the rest of the oscillators. The ground state will be the tensor product of the ground state of each such pair, together with the ground states of the unpaired oscillators at the boundary. The pairing is different for $N$ odd and even, so we will do these two cases separately. 

The reference state is defined as $\Psi_{k} |R\rangle=0$ which is the same thing as $\Psi_{n} |R\rangle=0$ for $n,k \in [0,N-1]$. This state has no entanglement between any two oscillators on the lattice. 

When $N$ is odd, rewriting the Hamiltonian in the paired form yields
\bea
H = \omega [\Psi^{\dagger}_{0} , \Psi_{0}] + \sum^{\frac{N-1}{2}}_{k =1} \Big( \omega [\Psi^{\dagger}_{k} , \Psi_{k}] + \omega [\Psi^{\dagger}_{N-k} , \Psi_{N-k}] + 2 \sin {\frac{2 \pi k}{N}}  \big(\Psi^{\dagger}_{k}\Psi^{\dagger}_{N-k} - \Psi_{k}\Psi_{N-k}\big) \Big). \nonumber \\ 
\eea
This can be written as
\bea 
H = \sum^{\frac{N-1}{2}}_{k =1}H_{k} +\omega [\Psi^{\dagger}_{0} , \Psi_{0}] 
\eea
where $H_{k}$ is the hamiltonian for the two oscillators $\Psi_{k}$ and $\Psi_{N-k}$
for $ k \in [1, \frac{N-1}{2}]$. Defining $b_{1}=\Psi_{k}$ and $b_{2}=\Psi_{N-k}$
\bea
H_{k} = \omega [b^{\dagger}_{1}, b_{1}] + \omega [b^{\dagger}_{2}, b_{2}] + 2 \ \sin{\frac{2 \pi k}{N}}\ 
\big( b^{\dagger}_{1}b^{\dagger}_{2}- 
b_{1}b_{2} \big),
\eea
After B-V transformation
\bea
\tilde{b_{1}} = \rho b_{1} + \eta b^{\dagger}_{2}, \qquad \tilde{b_{2}} = \eta b^{\dagger}_{1} - \rho b_{2}
\eea
Here
\bea
\rho &=& \frac{s}{\sqrt{2} \sqrt{\omega(\omega- \sqrt{s^{2}+ \omega^{2}})+ s^{2}}}, \\ \nonumber
\eta &=&  \frac{\sqrt{\omega(\omega- \sqrt{s^{2}+ \omega^{2}})+ s^{2}}}{\sqrt{2} \sqrt{s^{2}+ \omega^{2}}},
\eea
where $s=\sin{\frac{2 \pi k}{N}}$.
The ground state is defined by $\tilde{b_{i}}| \tilde{0}\tilde{0} \rangle =0$ for $i\in\{1,2\}$. Writing
\bea
| \tilde{0}\tilde{0} \rangle= \alpha_{ij} | i j \rangle
\eea
where now $i,j \in (0,1)$, using B-V transformations we can show that the ground state takes the form  
\bea
| \tilde{0}\tilde{0} \rangle = 
\alpha_{00} |00 \rangle + 
\alpha_{11} | 11 \rangle
\eea
with $\alpha_{01} =\alpha_{10} = 0$. Furthermore
\bea
\rho \alpha_{00} = \eta \alpha_{11}
\eea
and
\bea |\alpha^{2}_{00}|+ |\alpha^{2}_{11}|= 1.
\eea
So here also the $U(4) \rightarrow U(2)$ reduction (analogous to the minimal  oscillator toy model) happens. The transformation from reference state to target state for all $N$ oscillators together can therefore be viewed as an element of
\bea
U(2)^{\otimes \frac{N-1}{2}}. 
\eea
This is a tensor product of $\frac{N-1}{2}$ factors of $U(2)$  because $\frac{N-1}{2}$ pairs of oscillators are getting mixed at a time when $N$ is odd.

Similarly when $N$ is even the hamiltonian in paired form is
\bea
H = \omega [\Psi^{\dagger}_{0} , \Psi_{0}] + \omega [\Psi^{\dagger}_{\frac{N}{2}} , \Psi_{\frac{N}{2}}] + \sum^{\frac{N-2}{2}}_{k =1} H_{k}
\eea 
With the $H_{k}$ here being the same as in the odd case. The only difference is that in the even case there are $\frac{N-2}{2}$ pairs and so the transformation from reference state to target state for all $N$ oscillators can be taken to be in
\bea
U(2)^{\otimes \frac{N-2}{2}}.
\eea

\subsubsection{Neveu-Schwarz Boundary Condition}

Let us now impose a Neveu-Schwarz boundary condition
\bea
\Psi_{n+N}=-\Psi_{n}
\eea
on our Majorana lattice. 
The discrete Fourier transform for this boundary condition is
\bea
\Psi_{n} = \frac{1}{\sqrt{N}} \sum^{N-1}_{k =0} e^{\frac{2 \pi in(k + \frac{1}{2})}{N}} \Psi_{k}
\eea
and the inverse transform
\bea
\Psi_{k} = \frac{1}{\sqrt{N}} \sum^{N-1}_{n =0} e^{\frac{-2 \pi in(k + \frac{1}{2})}{N}} \Psi_{n}.
\eea
Although the $\Psi_{n}$ satisfies Neveu-Schwarz boundary condition the $\Psi_{k}$ satisfies an analogue of the Ramond boundary condition in Fourier space:
\bea
\Psi_{k+N} = \Psi_{k}.
\eea
The Hamiltonian takes the form 
\bea
H  &=& \sum^{N-1}_{n =0}  \big( \omega [\Psi^{\dagger}_{n} , \Psi_{n}] -i (\Psi_{n} \Psi_{n+1} +\Psi^{\dagger}_{n}\Psi^{\dagger}_{n+1}) \big)\\ \nonumber
&=&\sum^{N-1}_{k =0}  (\omega [\Psi^{\dagger}_{k} , \Psi_{k}] -i (\Psi_{k} \Psi_{-k-1}e^{\frac{-2 \pi i(k + \frac{1}{2})}{N}} +
\Psi^{\dagger}_{k}\Psi^{\dagger}_{-k-1} e^{\frac{2 \pi i(k+ \frac{1}{2})}{N}}))
\eea
Using periodicity in $k$, we can use
\bea
\Psi_{-k-1} = \Psi_{N-k-1}
\eea
to rewrite the Hamiltonian as
\bea
H = \sum^{N-1}_{k =0}  (\omega [\Psi^{\dagger}_{k} , \Psi_{k}] -i (\Psi_{k} \Psi_{N-k-1}e^{\frac{-2 \pi i(k + \frac{1}{2})}{N}} +
\Psi^{\dagger}_{k}\Psi^{\dagger}_{N-k-1} e^{\frac{2 \pi i(k+ \frac{1}{2})}{N}}))
\eea
When $N$ is even, the Hamiltonian in the paired form is
\bea
H &=& \sum^{\frac{N}{2}-1}_{k =0} (\omega [\Psi^{\dagger}_{k} , \Psi_{k}] + \omega [\Psi^{\dagger}_{N-1-k} , \Psi_{N-1-k}] + 
2 \sin \big[ {\frac{2 \pi}{N}(k +\frac{1}{2})} \big] (\Psi^{\dagger}_{k}\Psi^{\dagger}_{N-k-1} - \Psi_{k}\Psi_{N-k-1})) \nonumber \\
&\equiv &\sum^{\frac{N}{2}-1}_{k =0} H_{k}
\eea
As in the Ramond case, this can again be interpreted as a pair-wise mixing Hamiltonian, and the corresponding term of the Hamiltonian is $H_{k}$.
At each $k$, defining $b_{1}=\Psi_{k}$ and $b_{2}=\Psi_{N-1-k}$ (the index $k$ is suppressed in the operators $b$):
\bea
H_{k} = \omega [b^{\dagger}_{1} , b_{1}] + \omega [b^{\dagger}_{2} , b_{2}] + 2 \sin \big[ {\frac{2 \pi}{N}(k +\frac{1}{2})} \big]
(b^{\dagger}_{1}b^{\dagger}_{2} - b_{1}b_{2})
\eea
Comparing with the Ramond case, the only difference is that $k$ there is replaced by $k+\frac{1}{2}$ here. Following similar approach one can show here too that the $U(4) \rightarrow U(2)$ reduction happens. The transformation from the reference state to the target state for all $N$ oscillators is in
$U(2) ^{\otimes \frac{N}{2}}$.

Similarly when $N$ is odd the paired form of the Hamiltonian is 
\bea
H = \omega [\Psi^{\dagger}_{\frac{N-1}{2}} , \Psi_{\frac{N-1}{2}}] + \sum^{\frac{N-2}{2}}_{k =0} H_{k},
\eea
with the same $H_{k}$ as in the even case (for the Neveu-Schwarz boundary condition). This time there are $\frac{N-1}{2}$ pairs and so the transformation from the reference state to the target state for all $N$ oscillators is in $U(2)^{\otimes  \frac{N-1}{2}}$.

\section{Continuum Field Theory}

Everything we did so far was by discretizing spacetime into a lattice. This was the strategy adopted for the scalar case in \cite{Myers}, and what we have shown is that a similar strategy works (modulo minor --for our purposes-- subtleties like fermion doubling) for fermions as well. Now we will work directly in the continuum case following the corresponding approach for scalars undertaken in \cite{Chapman}. The idea here is basically a generalisation of the squeezing operator approach we discussed in passing in the lattice case. We can reach from an unentangled reference state $ |R \rangle $ to the entangled target state $ | \Psi \rangle $ via a unitary transformation
\bea
| \Psi \rangle = P e^{ -i \int_{\sigma_{i}}^{\sigma_{f}}{ G_{( \sigma)} d \sigma}} |R \rangle \label{P-unitary}
\eea
where in many cases, we will find that the $G(\sigma)$ can be realised via an appropriate squeezing operator, 
that creates quantum entanglement below some UV cut off scale $\Lambda$. Here $ \sigma $ parametrises our path such that at $ \sigma = \sigma_{i} $ we have our reference state $ |R \rangle $ and at $ \sigma= \sigma_{f} $ we get the target state $ | \Psi \rangle $. The path ordering $P$ is not required for commuting generators $ G(\sigma) $, as will often be the case if we manage to find appropriate squeezing operators. We will begin with Dirac fermions in 1+1 dimensions, which is a standard context where the cMERA tensor network is discussed \cite{cMERA, Takayanagi}, with an eye towards our discussions in the next section.

\subsection{Dirac in 1+1 Dimensions}

We consider the Dirac Hamiltonian in 1+1 dimensions given by
\bea
H = \int {dx[ -i\overline{\Psi} \gamma^{x} \partial_{x} \Psi + m \overline{\Psi} \Psi]}
\eea
Here $ \Psi = ( \Psi_{1} , \Psi_{2})^T $ is the two component complex fermion and $\gamma^{t}=\sigma_{3}$ and $\gamma^{x}=i \sigma_{2}$.\\
By a Fourier transform  $$\Psi_{i}(x)=\int \frac{dk}{\sqrt{2 \pi}} \Psi_{i}(k) e^{i k x}$$ we can write the hamiltonian as
\bea
H = \int {dk [k \Psi_{1}^{ \dagger } (k) \Psi_{2} (k) +
 k\Psi_{2}^{\dagger}(k) \Psi_{1} (k) + m\Psi_{1} ^{\dagger } (k) \Psi_{1}(k) - m\Psi_{2}^{\dagger }(k) \Psi_{2}(k)]}
\eea
The canonical anti-commutators in momentum space become
\bea
\{\Psi_{1} (k), \Psi_{1}^{\dagger } (k^{'})\}  = \{ \Psi_{2} (k) , \Psi_{2}^{\dagger} (k^{'}) \} = \delta (k-k') 
\eea 
The reference unentangled IR State $|R\rangle$ can be defined by
\bea
\Psi_{1} (k) | R \rangle = 0 , \: \: \Psi_{2} ^{\dagger} (k)| R \rangle = 0 \quad \forall k \in \R \label{Dirac-Ref}
\eea
This is the ground state of the ultra local Hamiltonian
\bea 
H_{m} = \int {dx ( m \overline{\Psi} \Psi)} =\int {dk\ m\ \big(\Psi_{1} ^{\dagger } (k) \Psi_{1}(k) - \Psi_{2}^{\dagger }(k) \Psi_{2}(k)\big)}
\eea
The ground state of this fermionic theory can be obtained by direct analogy with the discrete case via a Bogoliubov-Valatin transformation\footnote{Note that this same ground state is precisely the one that is obtained in the standard approach to quantization of Dirac fermions, which involves the introduction of Dirac $u$ and $v$ modes. We demonstrate this in appendix C.} which is 
\bea
\tilde{\Psi}_{1} (k) |\Psi \rangle = 0 , \: \: \tilde{\Psi}^{\dagger}_{2} (k) | \Psi \rangle = 0   \quad \forall k \in \R
\eea
where 
\bea
\tilde{\Psi}_{1} (k)&=&(A_{k} \Psi_{1} (k) + B_{k} \Psi_{2} (k) )\\ \nonumber
\tilde{\Psi}_{2} (k) &=&(-B_{k} \Psi_{1} (k) + A_{k} \Psi_{2} (k) )
\eea
are the normal mode coordinates and
\bea 
A_{k} = \frac{-k}{\sqrt{k^{2} +( \sqrt{ k^2 + m^2} - m)^{2} }}, \\ B_{k} = \frac{m- \sqrt{k^{2}+ m^{2} }}{ \sqrt{k^{2} +( \sqrt{ k^2 + m^2} - m)^{2}}}. 
\eea
Noting that $A_{k} ^{2} + B_{k} ^{2} =1$, we introduce 
\bea
r_{k} = -\frac{1}{2}\tan ^{-1} \left( \frac{k}{m} \right) \label{rk1+1}
\eea
so that 
\bea
A_{k} =- \cos (r_{k}) \ \  {\rm and} \ \ \; B_{k} =  \sin (r_{k})
\eea
Note the similarity of this expression $r_k$ with analogous expressions in the discrete case.

As our target state, following the scalar case in \cite{Chapman}, we consider the approximate ground state $ |m^{( \Lambda )} \rangle  $ which is defined as 
\bea
\tilde{\Psi}_{1} (k) |m^{( \Lambda )} \rangle = 0 , \: \: \tilde{\Psi}^{\dagger}_{2} (k) |m^{( \Lambda )} \rangle = 0 \quad \forall k: |k| \leq \Lambda \\ \nonumber
\Psi_{1} (k)|m^{( \Lambda )} \rangle= 0 , \: \: \Psi_{2} ^{\dagger} (k) |m^{( \Lambda )} \rangle= 0 \quad \forall k: |k| > \Lambda
\eea
We can reach target state from reference state via a unitary transformation of the form
\bea
|m^{ (\Lambda)} \rangle = e^{-i \int_{|k| \leq \Lambda} dk K(k)r_{k}} |R \rangle
\eea
where the reference state is $|R \rangle=|\Omega\rangle \quad \forall k$ 
and $K(k)$ is the squeezing operator which we define as
\bea
K(k) = i (\Psi_{1}^{\dagger}(k) \Psi_{2}(k) + \Psi_{1}(k) \Psi_{2}^{\dagger} (k)) \label{Dirac-squeeze}
\eea
We will see that this construction shares many of the features of the scalar case discussed in \cite{Chapman}, despite the fact that here the entanglement is happening between the two fermionic modes, and not between modes at antipodal momenta. Note also that the target ground state that we have defined is {\em not} the cMERA ground state \cite{cMERA, Takayanagi} even though in the $\Lambda \rightarrow \infty$ limit they both tend to the true ground state. The cMERA can be viewed as a non-geodesic path in our language, as we will discuss in the next section. 
 
\subsection{Ground State Complexity }

But before we do all that, let us evaluate the complexity using our squeezing operator above. With the specific choice of the squeezing operator we have made in the previous section, the calculation goes entirely parallel to the one in \cite{Chapman}, but we review it here largely to establish our notation. First, we calculate the distance with the Fubini-Study metric:
\bea
ds_{FS} (\sigma) = d \sigma \sqrt{|\partial_{\sigma}|\Psi(\sigma)\rangle|^{2} - |\langle \Psi(\sigma)|\partial_{\sigma}|\Psi(\sigma)|^{2}}
\eea
and then minimise the length. 
The circuit length of a path traced by intermediate states $ | \Psi (\sigma) \rangle = U( \sigma) | R \rangle $ is
\bea
l(| \Psi(\sigma) \rangle) = \int_{\sigma_{i}}^{\sigma_{f}} ds_{FS}(\sigma)
\eea
Using \eqref{P-unitary} we can write the FS metric as
\bea
ds_{FS} ( \sigma) = d \sigma \sqrt{\langle G^{2}(\sigma) \rangle _{ \Psi(\sigma)} - \langle G(\sigma) \rangle_{\Psi(\sigma)}^{2}}
\eea
 
In the context of our discussion in the previous subsection, we can consider a general path generated by the squeezing operator as a unitary transformation that takes the state through
\bea
U( \sigma)=e^{-i \int_{|k| \leq \Lambda} dk K_{(k)}Y_{k}(\sigma)} |R \rangle
\eea
Here 
\bea
Y_{k}(\sigma) = \int_{s_{i}}^{\sigma} y_{k} (s') ds' \quad { \rm and} \quad Y_{k}(s_{f}) = r_{k}
\eea
where the last condition arises from our specific choice of target state. 
The order of $k$ and $\sigma$ integrals can be exchanged because $K(k)$ at different $k$ are all commuting. By direct calculation one sees that
\bea
\langle G(\sigma) \rangle = 0, \ \ {\rm and} \ \  \langle G^{2}( \sigma) \rangle = {\rm Vol} \int dk\  y_{k}^{2} (\sigma).
\eea
Here we write $ \delta (0) = {\rm Vol} $. The complexity is the length of the minimal path:
\bea
C= \underset{ \{ G(s) \} } { { \rm min}} \quad l(ds_{FS})=
\underset{ \{ Y_{k}(\sigma) \} } {{ \rm min}} \int_{s_{i}}^{s_{f}} d\sigma \sqrt{ { \rm Vol} \int_{|k| \leq \Lambda} dk
 (\partial_{\sigma} Y_{k} (\sigma))^{2}} 
\eea 
We recognise a flat Euclidean geometry associated with coordinate $ Y_{k}(\sigma) $, implying that the geodesic is a straight line path. Under a simple affine parametrisation $\sigma $ the geodesic is
\bea
Y_{k}(\sigma) = \frac{ \sigma - s_{i}}{s_{f} - s_{i}} Y_{k}(s_{f})
\eea 
The complexity for the target state is the length of this straight line path
\bea
C = \sqrt{{ \rm Vol} \int_{k \leq \Lambda} dk (r_{k}^{2})}
\eea
and $r_{k}$ is given by \eqref{rk1+1}. As we see, the essential difference between complexity in the scalar case and the fermionic case is in the form of the $r_k$, both in the discrete case and the continuum case. We find that the parallel between the form of the complexity in the discrete \cite{Myers} and continuum \cite{Chapman} cases for the scalar, holds for the fermionic case as well.

These integrals are divergent in the cut-off $\Lambda$. In the massive case, the above expression can be explicitly evaluated to

\bea
\begin{split}
C^{2}=-\frac{m}{2} {\rm Vol} \log \left(\frac{\Lambda ^2}{m^2}+1\right) \tan ^{-1}\left(\frac{\Lambda }{m}\right)-\frac{1}{4} m {\rm Vol} \left(-\log (16) \tan ^{-1}\left(\frac{\Lambda }{m}\right)+ \right.\\\left. -i \text{Li}_2\left(-e^{-2 i \tan ^{-1}\left(\frac{\Lambda }{m}\right)}\right)+i \text{Li}_2\left(-e^{2 i \tan ^{-1}\left(\frac{\Lambda }{m}\right)}\right)\right)+\frac{1}{2} \Lambda  {\rm Vol} \left(\tan ^{-1}\left(\frac{\Lambda }{m}\right)\right)^2 
\end{split}
\eea
Its behaviour at large $\Lambda$ takes the form
\bea
C^{2}=\frac{\pi ^2}{8}  {\rm Vol}\left(\Lambda +\frac{4 m}{\pi}\log \left(\frac{2 m}{\Lambda }\right)-\frac{4 m}{\pi}+ {\cal O}( 1/\Lambda)\right)
\eea
When $m=0$ the integral is simple and the complexity simplifies:
\bea
C=\sqrt{\frac{\pi^{2}\Lambda{\rm Vol}}{8}}
\eea


\subsection{Ground State Complexity and $SU(2)$ Generators}

Our discussion so far used a single generator $K(k)$ to reach the target state. This can be viewed as a specific choice of gate in the circuit complexity language. Since we were able to construct a parallel between the entangling operator in the scalar case in \cite{Chapman} and the Dirac fermion case above, it is interesting to ask whether a construction analogous to the more general $SU(1,1)$ generators discussed in \cite{Chapman} is possible here. It turns out that the answer is yes, except that the generators satisfy an $SU(2)$ algebra now. Our calculations in this section are direct adaptations of those in the letter of \cite{Chapman}, but we include some tricks here that simplify the calculation. 

The basic idea is to note that the target state can also be reached by using a more general set of generators 
\bea
K_{+}(k) &=& i\Psi_{1}^{ \dagger} (k)\Psi_{2}(k) \\ \nonumber
K_{-}(k) &=& i\Psi_{1} (k)\Psi_{2}^{\dagger} (k)\\ \nonumber
K_{0}(k) &=& \frac{\Psi_{1}^{\dagger} (k)\Psi_{1} (k)- \Psi_{2}^{\dagger} (k)\Psi_{2}(k)}{2}
\eea
These commute with the number preserving operators $(n_{1} - n_{2})$ where $ n_{1} = \Psi_{1}^{\dagger} \Psi_{1}$ and $ n_{2} = \Psi_{2} \Psi_{2}^{\dagger} $.
These generators satisfy the following commutations relations
\bea
[ K_{+}(k), K_{-}(k')] &=& 2K_{0}(k)\ \delta(k-k^{'}) \\ \nonumber
[K_{0}(k) ,K_{+}(k')] &=&K_{+}(k)\  \delta(k-k^{'}) \\ \nonumber
[K_{0}(k) , K_{-}(k')] &=& -K_{-}(k)\ \delta(k-k^{'})
\eea
This is easily seen to be a set of decoupled $SU(2)$ algebras at each $k$, once one rescales the generators appropriately with $ \delta (0) = {\rm Vol} $. Now as in \cite{Chapman} let us consider a general path of the form
\bea
| \Psi(\sigma) \rangle = e^{  \int_{|k|< \Lambda}{dk \ g(k,\sigma)}} | R \rangle
\eea
generated by $g(k,\sigma)$, given as
\bea
g(k,\sigma) = \alpha_{+}(k,\sigma)K_{+}(k) + \alpha_{-}(k, \sigma)K_{-}(k) + \omega(k,\sigma)K_{0}(k).
\eea
Unitarity condition for above transformation implies $ \alpha_{+}^{*} (\sigma) = -\alpha_{-}(\sigma) $ and $ \omega^{*}(\sigma) = -\omega(\sigma) $ 
We can decompose the unitary transformation above using \cite{Book}
\bea
U(\sigma) = e^{\int_{|k|< \Lambda}{dk \beta_{+}(k,\sigma)K_{+}(k)}}e^{\int_{|k|< \Lambda}{dk  \ \log \beta_{0}(k, \sigma) K_{0}(k)}}e^{\int_{|k|< \Lambda}{dk \beta_{-}(k, \sigma)K_{-}(k)}} \label{g}
\eea
where the coefficients are 
\bea
\beta_{\pm} = \frac{2 \alpha_{\pm} \sinh \Xi}{2 \Xi \cosh\Xi - \omega \sinh \Xi}, \ \ 
\beta_{0} = (\cosh \Xi - \frac{\omega}{2 \Xi} \sinh \Xi)^{-2}. \label{beta-alpha} \nonumber
\eea
with $\Xi$ defined via
\bea
\Xi^{2} = \frac{\omega^{2}}{4} + \alpha_{+} \alpha_{-}
\eea
Analogous to \cite{Chapman} now we can observe that $K_{-} $ annihilates the reference state while exponential of $ K_{0} $ only changes the reference state upto a phase
\bea
K_{-}|R \rangle = 0 ; \quad K_{0}|R \rangle = -\frac{\delta(0)}{2} |R\rangle
\eea
So $| \Psi(\sigma) \rangle$ can be written as
\bea
| \Psi(\sigma) \rangle = \N e^{\int_{ \Lambda}{dk \beta_{+}(k, \sigma) K_{+}(k)}}|R \rangle
\eea
with 
\bea
\N = e^{\frac{-\delta(0)}{2}\int_{|k| \leq \Lambda} dk \ { \rm log} \beta_{0} (k, \sigma)}.
\eea

To compute the Fubini-Study metric from here is a bit of work, and we will introduce a small trick to accomplish it painlessly. Let us define
\bea
\tilde{K}_{+}=K_{-} \quad \tilde{K}_{-}=K_{+} \quad \tilde{K}_{0}=-K_{0} 
\eea
The crucial fact that makes them useful is that they satisfy identical commutation relations as the untilde'd operators:
\bea
[ \tilde{K}_{+}(k), \tilde{K}_{-}(k')] = 2\tilde{K}_{0}(k) \delta(k-k^{'}) \\ \nonumber
[\tilde{K}_{0}(k) ,\tilde{K}_{+}(k')] =\tilde{K}_{+}(k) \delta(k-k^{'}) \\ \nonumber
[\tilde{K}_{0}(k) , \tilde{K}_{-}(k')] = -\tilde{K}_{-}(k) \delta(k-k^{'})
\eea
The $g(k,\sigma)$ can be written as a linear combination of these generators as well
\bea
g(k,\sigma) = \tilde{\alpha}_{+}(k,\sigma)\tilde{K}_{+}(k) + \tilde{\alpha}_{-}(k, \sigma)\tilde{K}_{-}(k) + \tilde{\omega}(k,\sigma)\tilde{K}_{0}(k)
\eea
Comparing this with equation \eqref{g}, we get
\bea
\tilde{\alpha}_{+}=\alpha_{-} \quad \tilde{\alpha}_{-}=\alpha_{+} \quad \tilde{\omega}=-\omega \label{matching}
\eea
As the tilde'd generators {\em also} satisfy identical commutation relations, $U(\sigma)$ can be decomposed in the same fashion in terms of these generators as
\bea
U(\sigma) = e^{\int_{|k|< \Lambda}{dk \tilde{\beta}_{+}(k,\sigma)\tilde{K}_{+}(k)}}e^{\int_{|k|< \Lambda}{dk (log \tilde{\beta}_{0}(k, \sigma))\tilde{K}_{0}(k)}}e^{\int_{|k|< \Lambda}{dk \tilde{\beta}_{-}(k, \sigma)\tilde{K}_{-}(k)}}
\eea
where the coefficients satisfy exactly the same formulas as before, but now with tilde'd quantities. 
Using \eqref{matching} we can show that these are related to the coefficients in \eqref{beta-alpha} by 
\bea
\tilde{\beta^{*}_{\pm}} =- \beta_{\pm} \quad \tilde{\beta^{*}_{0}} = \beta_{0}
\eea
It is useful to write the decomposition of $U^{\dagger}(\sigma)$ with the latter generators
\bea
\begin{split}
U^{\dagger}(\sigma) = e^{\int_{|k|< \Lambda}{dk \tilde{\beta^{*}_{-}}(k,\sigma)\tilde{K}_{+}(k)}}e^{\int_{|k|< \Lambda}{dk (log \tilde{\beta^{*}_{0}}(k, \sigma))\tilde{K}_{0}(k)}}e^{\int_{|k|< \Lambda}{dk \tilde{\beta^{*}_{+}}(k, \sigma)\tilde{K}_{-}(k)}}\\
=e^{-\int_{|k|< \Lambda}{dk \beta_{-}(k,\sigma)K_{-}(k)}}e^{-\int_{|k|< \Lambda}{dk (log \beta_{0}(k, \sigma))K_{0}(k)}}e^{-\int_{|k|< \Lambda}{dk \beta_{+}(k, \sigma)K_{+}(k)}}
\end{split}
\eea
This last form helps us in substantially reducing the mindless labor involved in the calculations here as well as in the analogous results in \cite{Chapman}.

Further in this section we suppress some notations: integral over $k$ is just denoted with the integral symbol, and the argument $k$ is often not explicitly written. To calculate the Fubini-Study metric, we have
\bea
\partial_{\sigma} |\Psi(\sigma)\rangle=\Big(-\frac{{ \rm Vol}}{2} \int \frac{\beta_{0}'}{\beta_{0}}+ \int \beta_{+}' K_{+} \Big)| \Psi(\sigma) \rangle.
\eea
This leads to
\bea
\langle \Psi(\sigma) | \partial_{\sigma}| \Psi(\sigma) \rangle  \nonumber &=& -\frac{{ \rm Vol}}{2} \int \frac{\beta_{0}'}{\beta_{0}} + \int \beta_{+}'\langle \Psi(\sigma)|K_{+}| \Psi(\sigma) \rangle \\ \nonumber &=& -\frac{{ \rm Vol}}{2}  \int \frac{\beta_{0}'}{\beta_{0}} + \int \beta_{+}' \zeta_{+i} \langle R | K_{i}| R \rangle
 \\ \nonumber &=& \frac{- { \rm Vol}}{2} \Big( \int \frac{\beta_{0}'}{\beta_{0}} + \int \beta_{+}'\zeta_{+0} \Big)
\eea
where $\zeta_{ij}$ is defined via
\bea
U^{\dagger}K_{i}U = \zeta_{ij}K_{j}.
\eea
This leads to
\bea
|\langle \Psi(\sigma) | \partial_{\sigma}| \Psi(\sigma) \rangle|^{2}=\Big(\frac{ { \rm Vol}}{2}\Big)^{2} \Big( \int \frac{\beta'^{*}_{0}}{\beta^{*}_{0}} + \int \beta'^{*}_{+}\zeta^{*}_{+0} \Big) \Big( \int \frac{\beta_{0}'}{\beta_{0}} + \int \beta_{+}'\zeta_{+0} \Big)
\eea
It is useful to note that
\bea
\zeta^{*}_{+0}=\zeta_{-0}
\eea
The second piece is the Fubini-Study metric follows from a similar, but slightly lengthier calculation:
\bea
|\partial_{\sigma} |\Psi(\sigma)\rangle|^{2}=\Big(\frac{ { \rm Vol}}{2}\Big)^{2} \Big( \int \frac{\beta'^{*}_{0}}{\beta^{*}_{0}} + \int \beta'^{*}_{+}\zeta^{*}_{+0} \Big) \Big( \int \frac{\beta_{0}'}{\beta_{0}} + \int \beta_{+}'\zeta_{+0} \Big)
+{ \rm Vol} \int \beta_{+}^{*'} \beta_{+}^{'}\ \zeta_{--}\zeta_{++}\nonumber \\
\eea
So the final form of the metric is
\bea
ds_{FS}=d\sigma \sqrt{{ \rm Vol} \int \beta_{+}^{*'} \beta_{+}^{'} \zeta_{--}\zeta_{++}}
\eea
where
\bea
\zeta_{--}\zeta_{++} = \Big( 1 + \frac{\beta_{+}\beta_{-}}{\beta_{0}^{*}} \Big)^{2}=\frac{1}{|\beta_{0}|^{2}}
\eea
Making use of the identity $ |\beta_{0}| = 1+ |\beta_{+}|^{2} $, the  metric finally simplifies to 
\bea
ds_{FS} = d \sigma \sqrt{{ \rm Vol} \int_{|k| \leq\Lambda} dk \frac{\beta_{+}^{*'}\beta_{+}^{'}}{(1 + |\beta_{+}|^{2})^{2}}}
\eea
where the prime denote the derivative with respect to $\sigma$.
Thus the metric has the form of the Fubini-Study metric on $S^2$ ka $\IC\IP^1$ and the complexity arises as its geodesic.

\subsection{Another Generator}

Now that we have the $SU(2)$ generators in our hand, we can construct another lone generator $B(k)$ out of them that is distinct from $K(k)$, which takes us from the initial reference state to the target state, just like $K(k)$ does \cite{Chapman}. This takes the form
\bea
B(k) = -2 i \sin (r_{k})[K_{+} - K_{-}] -4 \cos (r_{k}) K_{0}
\eea
The unitary transformation which does the job is
\bea
|m^{(\Lambda)} \rangle = e^{i \frac{\pi}{4} \int_{|k|\leq \Lambda} dk B(k)} |R \rangle
\eea
Let the intermediate state be
\bea
|\Psi(\sigma)\rangle=e^{i \frac{\pi}{4} \int_{|k|\leq \Lambda} dk B(k) Y_{k}(\sigma)} |R \rangle
\eea
Plugging this in the fubini-study metric gives
\bea
ds_{FS}=d\sigma \frac{\pi}{2} \sqrt{{\rm Vol} \int_{|k|\leq \Lambda} dk \sin^{2}r_{k}  (\partial_{\sigma} Y_{k}(\sigma))^{2}}
\eea
So the path with the least length is
\bea
Y_{k}(\sigma)=\sigma.
\eea

We can compare the $B(k)$ form with the $SU(2)$ generators of the last section, and read off
\bea
\alpha_{+} = \frac{\pi \sigma}{2} \sin{r_{k}}, \quad \alpha_{-}= -\frac{\pi \sigma}{2} \sin{r_{k}}, \quad \omega = - i \pi \sigma \cos{r_{k}},  \quad \Xi = \frac{i \pi \sigma}{2},
\eea
which translates to the $\beta_{+}$:
\bea
\beta_{+} = \frac{i\sin{r_{k}} \sin{\frac{\pi \sigma}{2}}}{i\cos{\frac{\pi \sigma}{2}} - \cos{r_{k}} \sin{\frac{\pi \sigma}{2}}}. \label{B-beta}
\eea
At $\sigma =1 $ the $\beta_{+}$ for $B(k)$ becomes
\bea
\beta_{+} = -i\tan{r_{k}}.
\eea 
The length of this path, on plugging the expression \eqref{B-beta} into the $\IC \IP^1$ metric from the end of last section yields
\bea
l = \frac{\pi}{2}  \sqrt{{ \rm Vol} \int_{|k|\leq\Lambda}{ dk \sin^{2}{r}_{k}}}.
\eea
This can be explicitly evaluated to be
\bea
l^{2}=\left(\frac{\pi }{2}\right)^2 {\rm Vol} \left(\Lambda +\frac{1}{2} m \log \left(\frac{\sqrt{\Lambda ^2+m^2}-\Lambda }{\sqrt{\Lambda ^2+m^2}+\Lambda}\right)\right),
\eea
which goes as 
\bea
l^{2}=\frac{1}{4} \pi ^2  {\rm Vol}\ \Big(\Lambda+ m \log \frac{m}{2 \Lambda}+...\Big)
\eea
for large $\Lambda$, with the dots denoting sub-leading powers in $\Lambda$.

For comparison, for the straight line path using squeezing operator $K(k)$ the $SU(2)$ coefficients are 
\bea
\alpha_{+} = -i r_{k} \sigma;\quad \alpha_{-} = -i r_{k} \sigma ; \quad \omega = 0 ; \quad
\Xi = ir_{k} \sigma 
\eea
which gives the $\beta_{+}$ for $K(k)$ as
\bea
\beta_{+} = -i \tan({r_{k} \sigma}) \label{straightbeta}
\eea
The complexity of this path has been discussed in a previous subsection.

In an appendix, we explicitly show that the minimal path when the squeezing operator is $K(k)$ is a geodesic of the $\IC \IP^1$ metric, but the minimal path of the $B(k)$ operator is not.

\subsection{ Majorana Fermions in 1+1 Dimensions}

Now we move on to  field theories other than the 1+1 d Dirac fermion, which as we demonstrated, has very close parallels to the scalar field theory discussed in \cite{Chapman}. First we turn to the Majorana theory.

Previously we discussed the discrete version of 1+1 dimensional Majorana field theory. Now we consider the continuum version of it. The Lagrangian is
\bea
L = \int dx \ \overline{\psi} (i \gamma^{\mu}\partial_{\mu} -m) \psi 
\eea
where  the gamma matrices are the same ones we used in section 2, and
 \bea \gamma^{\mu}\partial_{\mu} =  \rho^{0} \partial_{0} + \rho^{1} \partial_{1}.
 \eea
The field $\psi$ is a 2 component spinor with its components real Grassmann variables classically:
\bea
\psi =
\left[ \begin{array} {c}
\Psi^{1} \\ \Psi^{2}
\end{array} \right]
\eea
The Lagrangian in terms of them becomes: 
\bea
L =\int dx \Big( i(\Psi^{1} \partial_{0}\Psi^{1} + \Psi^{2} \partial_{0} \Psi^{2} + \Psi^{1} \partial_{1} \Psi^{1} - \Psi^{2} \partial_{1} \Psi^{2}) +m \Psi^{1} \Psi^{2} -m \Psi^{2}\Psi^{1} \Big)
\eea
We developed our B-V technology more directly in the language of complex Grassmann variables, so we rewrite the Lagrangian in terms of
\bea
\Psi = \frac{\Psi^{1} - i \Psi^{2}}{\sqrt{2}} \quad , \quad \Psi^{\dagger} = \frac{\Psi^{1} + i \Psi^{2}}{\sqrt{2}}.
\eea
Here $\Psi , \Psi^{\dagger}$ are complex Grassmann variables. The Lagrangian becomes
\bea
L = \int dx \Big ( i (\Psi^{\dagger} \partial_{0} \Psi +\Psi \partial_{0} \Psi^{\dagger}+ \Psi \partial_{1} \Psi + \Psi^{\dagger} \partial_{1} \Psi^{\dagger}) -m [\Psi^{\dagger} , \Psi] \Big)
\eea
and the Hamiltonian
\bea
H = \int dx \Big( -i\Psi \partial_{1} \Psi -i\Psi^{\dagger} \partial_{1} \Psi^{\dagger} + m[\Psi^{\dagger} , \Psi] \Big).
\eea
Quantization proceeds by imposing the canonical anti-commutation relations
\bea
\{\Psi(x), \Psi^{\dagger}(x')\} = \delta(x-x')
\eea
and all other anti-commutators are zero. Doing a Fourier transform
\bea
\Psi(x) = \int \frac{dk}{\sqrt{2 \pi}} \Psi(k) e^{ikx}
\eea
they turn to
\bea
\{\Psi(k), \Psi^{\dagger}(k') \} = \delta(k-k')
\eea
and all other anti-commutators are zero. The Hamiltonian in Fourier variables is 
\bea
H=\int dk \big(-k \Psi(k) \Psi(-k) + k \Psi^{\dagger}(k) \Psi^{\dagger}(-k) + m[\Psi^{\dagger}(k),\Psi(k)]\big)
\eea
We define the reference state to be
\bea
\Psi(k)| R \rangle = 0
\eea
which is same as defining $ \Psi(x)| R \rangle = 0 $.
This state has no entanglement in position space and is also the ground state of the ultra-local Hamiltonian, ie., the above Hamiltonian without the terms arising from the derivatives in position space. 

In the ground state the oscillators at $k$ and $-k$ are to be entangled. The target state can be reached by using the squeezing operator $K(k)$ defined here as
\bea
K(k) = i(\Psi^{\dagger}(k) \Psi^{\dagger}(-k) + \Psi(k) \Psi(-k))
\eea
This operator entangles the oscillators at $k$ and $-k$.
Lets rewrite the Hamiltonian as
\bea
H = \int ^{\infty}_{0} dk \Big(    m[\Psi^{\dagger}(k) ,\Psi(k)] +m[\Psi^{\dagger}(-k) ,\Psi(-k)] + 2k (\Psi^{\dagger}(k)
\Psi^{\dagger}(-k) - \Psi(k) \Psi(-k)) \Big) \nonumber \\
\eea
In the Dirac case the integration limits ran from $-\infty$ to $\infty$ but here it is from $0$ to $\infty$. In other words, we have the freedom to define 
\bea
\Psi_1(k) \equiv \Psi(k), \ \ \Psi_2(k) \equiv \Psi^\dagger(-k)
\eea
here, and when we do it, we get the Dirac Hamiltonian that we wrote down earlier but with the integration over $k$ limited to the positive $k$ real axis. In other words, we are re-interpreting the entanglement between $k$ and $-k$ modes here as entanglement between two different fields, but at same $k$ (restricted to be positive). Since much of the calculations in the Dirac case goes through at each $k$ separately, this means that the Majorana calculation reduces loosely to half of Dirac. 
Apart from the restriction in the range of $k$, the ground states for both the cases are the same and the reference states are also the same. So the $r_{k}$ is the same as before and every derivation here is the same as before with the only difference being in the lower integration limit which is $0$ for this case and not $-\Lambda$ as in previous case. The complexity for the ground state of the Majorana theory is then
\bea
C = \sqrt{{\rm Vol} \int^{\Lambda}_{0} dk (r^{2}_{k})}
\eea
As $(r_{k})^{2}$ is an even function in $k$, we have
\bea
C_{Dirac}= \sqrt{2}  C_{Majorana}
\eea
where $C_{Dirac}$ and $C_{Majorana}$ are the complexities for the ground state of Dirac and Majorana theories in 1+1 dimensions.

\subsection{Fermions in Higher Dimensions}

Now we will generalize some of the above considerations to higher dimensions. The fermionic field theoretic hamiltonian in $d+1$ space-time dimensions is
\bea
H = \int d^{d}x \  \overline{\Psi}(x) (-i  \gamma^{i}\partial_{i} + m ) \Psi(x) 
\eea
where $i$ runs from $1$ to $d$. 
After doing a Fourier transform on each of the components of the spinor
\bea
\Psi_{a}(x) = \int \frac{d^{d}k }{(2 \pi)^{\frac{d}{2}}} \Psi_{a}(k) e^{i \vec{k}.\vec{x}}
\eea
The hamiltonian in Fourier variables is
\bea
H = \int d^{d} k \big( \overline{\Psi}(k) (\gamma^{i} k_{i} + m) \Psi(k) \big)
\eea

\subsubsection{Massless Theory in 3+1 Dimensions}
 
Let us start by considering a massless theory in 3+1 dimensions in chiral (a.k.a Weyl) basis:
\bea
\gamma^{0}=\left(
\begin{array}{cc}
 0 & I_{2} \\
 I_{2} & 0 \\
\end{array}
\right)
\quad
\gamma^{k}=\left(
\begin{array}{cc}
 0 & \sigma^{k} \\
 -\sigma^{k} & 0 \\
\end{array}
\right)
\eea
For massless particles the two Weyl spinors decouple from each other. The Hamiltonian is
\bea
H = \int d^{3} k \big(  \overline{\Psi}(k) \gamma^{i}k_{i} \Psi(k) \big)=\int d^{3} k \mathcal{H}(k)
\eea
where $i$ runs from 1 to 3. Each component of the spinor $\Psi(k)$ can be interpreted as an oscillator at $k$.
At each $k$ there are four fermionic oscillators with the Hamiltonian $\mathcal{H}(k)$. 
\bea
\mathcal{H}(k)= \frac{k_{3}}{2}[\Psi_{1}, \Psi^{\dagger}_{1}] + \frac{k_{3}}{2}[\Psi^{\dagger}_{2}, \Psi_{2}] + \frac{k_{3}}{2}[\Psi^{\dagger}_{3}, \Psi_{3}] + \frac{k_{3}}{2}[\Psi_{4}, \Psi^{\dagger}_{4}]+ \hspace{1.5in} \\ \nonumber \hspace{1in}
- (k_{1}- i k_{2}) \Psi^{\dagger}_{1} \Psi_{2} - (k_{1}+i k_{2}) \Psi^{\dagger}_{2} \Psi_{1} + (k_{1}-i k_{2}) \Psi^{\dagger}_{3} \Psi_{4} + (k_{1}+i k_{2}) \Psi^{\dagger}_{4} \Psi_{3}
\eea
In the above equation the dependence of the fields on $k$ is suppressed. The two oscillators $\Psi_{1}(k)$ and $\Psi_{2}(k)$ are decoupled from $\Psi_{3}(k)$ and $\Psi_{4}(k)$.

For each $k$, let us define $a_{1} = \Psi^{\dagger}_{1} , a_{2} = \Psi_{2} , b_{1} = \Psi_{3}, b_{2} = \Psi^{\dagger}_{4}$ so that
\bea
\mathcal{H}(k)=\mathcal{H}^{a}(k)+\mathcal{H}^{b}(k)
\eea
with 
\bea
\mathcal{H}^{a}(k)&=& \frac{k_{3}}{2}[a^{\dagger}_{1},a_{1}] + \frac{k_{3}}{2}[a^{\dagger}_{2},a_{2}] - (k_{1}-ik_{2}) a_{1}a_{2} +  (k_{1}+ik_{2}) a^{\dagger}_{1}a^{\dagger}_{2}   \\ \nonumber
\mathcal{H}^{b}(k)&=& \frac{k_{3}}{2}[b^{\dagger}_{1},b_{1}] + \frac{k_{3}}{2}[b^{\dagger}_{2},b_{2}]+  (k_{1}-ik_{2}) b^{\dagger}_{1}b^{\dagger}_{2} -(k_{1}+ik_{2}) b_{1}b_{2}
\eea

After the by-now-familiar B-V transformation we get
\bea
\mathcal{H}(k) = \frac{\omega_{k}}{2} \sum^{2}_{i =1}([\tilde{a}^{\dagger}_{i}, \tilde{a_{i}}] +[\tilde{b}^{\dagger}_{i}, \tilde{b_{i}}] )
\eea
where $ \omega_{k} = \sqrt{k^{2}_{1} +k^{2}_{2}+ k^{2}_{3} }$. 
The B-V transformation for $a$ type oscillators is
\bea
\left(
\begin{array}{c}
 \tilde{a}_{1} \\
 \tilde{a}_{2} \\
 \tilde{a}^{\dagger}_{1} \\
 \tilde{a}^{\dagger}_{2} \\
\end{array}
\right)= \sqrt{\frac{1}{2}\Big(1 -\frac{k_{3}}{\omega_{k}}\Big)}
\left(
\begin{array}{c}
a^{\dagger}_{2} +a_{1} \frac{\left(k_{3}+\omega_{k}\right)}{(k_{1}+i k_{2})}\\
 a^{\dagger}_{1}-a_{2} \frac{\left(k_{3}+ \omega_{k}\right)}{(k_{1}+i k_{2})} \\
 a_{2} +a^{\dagger}_{1} \frac{\left(k_{3}+\omega_{k}\right)}{(k_{1}-i k_{2})} \\
a_{1}- a^{\dagger}_{2} \frac{ \left(k_{3}+\omega_{k}\right)}{(k_{1}-i k_{2})} \\
\end{array}
\right)
\eea
and the B-V transformation for $b$ type oscillators is
\bea
\left(
\begin{array}{c}
 \tilde{b}_{1} \\
 \tilde{b}_{2} \\
 \tilde{b}^{\dagger}_{1} \\
 \tilde{b}^{\dagger}_{2} \\
\end{array}
\right)=\sqrt{\frac{1}{2}\Big(1 -\frac{k_{3}}{\omega_{k}}\Big)} \left(
\begin{array}{c}
 b^{\dagger}_{2} +b_{1} \frac{\left(k_{3}+\omega_{k}\right)}{(k_{1}-i k_{2})}  \\
b^{\dagger}_{1}-b_{2}  \frac{\left(k_{3}+\omega_{k}\right)}{(k_{1}-i k_{2})}  \\
b_{2} +b^{\dagger}_{1}  \frac{\left(k_{3}+\omega_{k}\right)}{(k_{1}+i k_{2})} \\
 b_{1} -b^{\dagger}_{2}\frac{ \left(k_{3}+\omega_{k}\right)}{(k_{1}+i k_{2})}\\
\end{array}
\right).
\eea

Here there is no natural ultra local Hamiltonian, because mass is zero. We can choose any reference state which has no entanglement in the physical (ie., $x$) space. Let's define the reference state $|R\rangle$ to be the state which is annihilated by $\Psi_{1}(x)$ , $\Psi^{\dagger}_{2}(x)$ , $\Psi^{\dagger}_{3}(x)$ and $\Psi_{4}(x)$ $\forall x \in \R^{3}$. This is the same thing as defining the reference state to be annihilated by $a_{i}(k)$ and $b_{i}(k)$ for $i=1,2$ and $\forall k \in \R^{3}$.

We define the target state to be the approximate ground state $|T^{(\Lambda)}\rangle$ defined as
\bea
\tilde{a}_{i}(k)|T^{(\Lambda)}\rangle=0  \quad \tilde{b}_{i}(k)|T^{(\Lambda)}\rangle=0 \qquad \forall k : |k| \leq \Lambda \\ \nonumber
a^{\dagger}_{i}(k)|T^{(\Lambda)}\rangle=0  \quad b^{\dagger}_{i}(k)|T^{(\Lambda)}\rangle=0 \qquad \forall k : |k| > \Lambda
\eea
It is possible to see that the target state can be reached from the reference state by the unitary transformation
\bea
|T^{(\Lambda)}\rangle=e^{-i \int_{|k| \leq \Lambda} d^{3}k \  r(k) K(k)}| R \rangle
\eea
where the squeezing operator $K(k)$ here is 
\bea
K(k) = i  (k_{1}- ik_{2}) (\tilde{a}^{\dagger}_{1}(k) \tilde{a}^{\dagger}_{2}(k)+\tilde{b}_{1}(k) \tilde{b}_{2}(k) )+ 
 i(k_{1}+ ik_{2}) (\tilde{a}_{1}(k) \tilde{a}_{2}(k)+\tilde{b}^{\dagger}_{1}(k) \tilde{b}^{\dagger}_{2}(k) ) \\ \nonumber
 =i  (k_{1}- ik_{2}) (a_{2}(k) a_{1}(k)+b^{\dagger}_{2}(k) b^{\dagger}_{1}(k) )+ 
 i(k_{1}+ ik_{2}) (a^{\dagger}_{2}(k)a^{\dagger}_{1}(k)+b_{2}(k) b_{1}(k) ) 
\eea 
and $r(k)$ is
\bea
r(k)=-\frac{1}{\sqrt{k^{2}_{1} + k^{2}_{2}}} \arctan{\Big( \frac{k_{3}+\omega_{k}}{\sqrt{k^{2}_{1} + k^{2}_{2}}} \Big)}
\eea
It can also be checked that the unitary transformation takes the $a_{i}$'s and $b_{i}$'s to  $\tilde{a}_{i}$'s and $\tilde{b}_{i}$'s via the similarity transformations
\bea
Ua^{\dagger}_{2}U^{\dagger}=\tilde{a}_{1} \qquad
Ua^{\dagger}_{1}U^{\dagger}=\tilde{a}_{2}\\ \nonumber
Ub^{\dagger}_{2}U^{\dagger}=\tilde{b}_{1}\qquad
Ub^{\dagger}_{1}U^{\dagger}=\tilde{b}_{2}.
\eea
As the annihilation operators of the reference state are transformed to the annihilation operators of the target state via the similarity transformation, the unitary transformation takes the reference state to the target state.

Now let's consider an arbitrary path generated by the squeezing operator
\bea
|\Psi(\sigma)\rangle=e^{-i \int^{\sigma}_{0} ds \int_{|k| \leq \Lambda} d^{3}k \   y_{k}(s) K(k)}| R \rangle \\ \nonumber
=e^{-i \int_{|k| \leq \Lambda} d^{3}k \ Y_{k}(\sigma) K(k)}| R \rangle
\eea
Note that by identifying a judicious squeezing operator, here too we have bypassed the need to do any path ordering because all $K(k)$ commute.

Evaluating the length of this path using Fubini-Study metric gives 
\bea
l(|\Psi(\sigma)\rangle)=\ \int_{0}^{1} d\sigma \sqrt{2 { \rm Vol } \int_{|k| \leq \Lambda} d^{3}k (k^{2}_{1} + k^{2}_{2})
 (\partial_{\sigma} Y_{k} (\sigma))^{2}} 
\eea
Doing a "coordinate" transformation
\bea
X_{k}(\sigma)=\sqrt{k^{2}_{1} + k^{2}_{2}}  Y_{k} (\sigma)
\eea
the length of the path becomes
\bea
l(|\Psi(\sigma)\rangle)=\ \int_{0}^{1} d\sigma \sqrt{2 { \rm Vol} \int_{|k| \leq \Lambda} d^{3}k 
 (\partial_{\sigma} X_{k} (\sigma))^{2}} 
\eea
This is again a flat Euclidean geometry associated with coordinate $X_{k} (\sigma)$ and so the geodesic is
\bea
X_{k}(\sigma)=  \sqrt{k^{2}_{1} + k^{2}_{2}}\ r(k) \  \sigma 
\eea
The complexity is the length of this path
\bea
C = \sqrt{ 2 {\rm Vol} \int_{|k|\leq \Lambda} d^{3}k \Bigg( \arctan  \Big( \frac {k_{3}+\omega_{k}}{\sqrt{k^{2}_{1} + k^{2}_{2}}} \Big) \Bigg)^{2} }
\eea
This integral can in principle be explicitly evaluated, but we will not present it here. Instead we go on to the massive case, where we will present all the gory details.

\subsubsection{Ground State Complexity and $SU(2) \times SU(2)$ Generators}

Let us consider an extended set of paths as we did before in section 3.3. We introduce $SU(2) \times SU(2)$ generators and consider the paths generated by them. More general paths are possible as we will see in the next subsection, but this is a sufficiently interesting generalisation that contains our squeezing operator. The squeezing operator used in the previous subsection can be written as
\bea
K(k) = i (k_{1}- ik_{2}) (\tilde{a}^{\dagger}_{1}(k) \tilde{a}^{\dagger}_{2}(k)+\tilde{b}_{1}(k) \tilde{b}_{2}(k) )+ 
 i(k_{1}+ ik_{2}) (\tilde{a}_{1}(k) \tilde{a}_{2}(k)+\tilde{b}^{\dagger}_{1}(k) \tilde{b}^{\dagger}_{2}(k) )\\ \nonumber
 =i (k_{1}- ik_{2}) (a_{2}(k) a_{1}(k)+b^{\dagger}_{2}(k) b^{\dagger}_{1}(k) )+ 
 i(k_{1}+ ik_{2}) (a^{\dagger}_{2}(k)a^{\dagger}_{1}(k)+b_{2}(k) b_{1}(k) ) 
\eea 
Let $(k_{1}+ ik_{2})=\kappa e^{i \theta}$ and $(k_{1}- ik_{2})=\kappa e^{-i \theta}$ and by absorbing the phases into the ladder operators the squeezing operator simplifies to
\bea
K(k) &=& i \kappa   (\tilde{a}^{\dagger}_{1}(k) \tilde{a}^{\dagger}_{2}(k)+\tilde{a}_{1}(k) \tilde{a}_{2}(k))
 +i \kappa (\tilde{b}^{\dagger}_{1}(k) \tilde{b}^{\dagger}_{2}(k) +\tilde{b}_{1}(k) \tilde{b}_{2}(k)  ) \\ \nonumber
 &=&i \kappa   (a^{\dagger}_{2}(k)a^{\dagger}_{1}(k) +a_{2}(k) a_{1}(k)) 
  + i \kappa (b^{\dagger}_{2}(k) b^{\dagger}_{1}(k)+b_{2}(k) b_{1}(k) )  \\ \nonumber
  &=&\kappa K^{a}(k)+ \kappa K^{b}(k)
\eea
where $K^{a}(k)\equiv i(a^{\dagger}_{2}(k)a^{\dagger}_{1}(k) +a_{2}(k) a_{1}(k))$ and $K^{b}(k)\equiv i(b^{\dagger}_{2}(k) b^{\dagger}_{1}(k)+b_{2}(k) b_{1}(k) ) $.

Notice from the Hamiltonian in the previous section the $a$ type oscillators and $b$ type oscillators completely decouple from each other. We can introduce the following generators
\bea
\begin{split}
&K^{a}_{+}(k)=  i a^{\dagger}_{2}(k)a^{\dagger}_{1}(k)&                          \quad &,& \quad             & K^{b}_{+}(k)=  i b^{\dagger}_{2}(k)b^{\dagger}_{1}(k)&\\
&K^{a}_{-}(k)= i  a_{2}(k) a_{1}(k)&                                                           \quad &,&\quad             &K^{b}_{-}(k)= i  b_{2}(k) b_{1}(k)&\\ 
&K^{a}_{0}(k)=\frac{a^{\dagger}_{2}(k)a_{2}(k)-a_{1}(k)a^{\dagger}_{1}(k)}{2}&       \quad &,&\quad             &K^{b}_{0}(k)=\frac{b^{\dagger}_{2}(k)b_{2}(k)-b_{1}(k)b^{\dagger}_{1}(k)}{2}&\\
\end{split}
\eea
Together these 6 generators generate a $SU(2) \times SU(2)$ algebra. Each of these generators commute with the number preserving operators $(n_{1}-n_{2})$ and $(n_{3}-n_{4})$ where 
\bea
n_{1}=a^{\dagger}_{2}a_{2}  ,\quad  n_{2}=a^{\dagger}_{1}a_{1} ,\quad n_{3}=b^{\dagger}_{2}b_{2}  ,\quad  n_{4}=b^{\dagger}_{1}b_{1}.
\eea
Let us consider the path generated by an arbitrary linear combination of these 6 generators
\bea
|\Psi(\sigma)\rangle=e^{\int_{|k|\leq \Lambda} dk g(k,\sigma)}|R\rangle
\eea
where $|R\rangle$ is the reference state defined in previous section and here $g(k,\sigma)\equiv g^{a}(k,\sigma)+g^{b}(k,\sigma)$ and
\bea
\begin{split}
g^{a}(k,\sigma)=\alpha^{a}_{+}(k,\sigma)K^{a}_{+}(k) + \alpha^{a}_{-}(k, \sigma)K^{a}_{-}(k) + \omega^{a}(k,\sigma)K^{a}_{0}(k) \\ 
g^{b}(k,\sigma)=\alpha^{b}_{+}(k,\sigma)K^{b}_{+}(k) + \alpha^{b}_{-}(k, \sigma)K^{b}_{-}(k) + \omega^{b}(k,\sigma)K^{b}_{0}(k)
\end{split}
\eea
The $K^{a}_{i}$'s commute with $K^{b}_{j}$'s (for $i$,$j=+,-,0$), so
\bea
e^{\int_{|k|\leq \Lambda} dk g(k,\sigma)}=e^{\int_{|k|\leq \Lambda} dk g^{a}(k,\sigma)}e^{\int_{|k|\leq \Lambda} dk g^{b}(k,\sigma)}
\eea
Proceeding in the same way as in section 3.3 we get
\bea
|\Psi(\sigma)\rangle=\N^{a} \N^{b} e^{\int_{ \Lambda}{dk \  \beta^{a}_{+}(k, \sigma) K^{a}_{+}(k)}} e^{\int_{ \Lambda}{dk \ \beta^{b}_{+}(k, \sigma) K^{b}_{+}(k)}} |R\rangle
\eea
with
\bea
\N^{a} = e^{\frac{-\delta(0)}{2}\int_{|k| \leq \Lambda} dk \ { \rm log} \beta^{a}_{0} (k, \sigma)} \qquad \N^{b} = e^{\frac{-\delta(0)}{2}\int_{|k| \leq \Lambda} dk \ { \rm log} \beta^{b}_{0} (k, \sigma)}
\eea
and the $\beta^{a}_{i}$'s and $\beta^{b}_{i}$'s are defined exactly analogous to the 1+1 d case.

Computing the Fubini-Study metric will give two copies of the metric on $S^{2}$ and the length of the whole path is minimised if the length of each individual paths (i.e $\beta^{a}_{+}$ and $\beta^{b}_{+}$) are minimised. As these are just two copies of what we had before in section 3, we conclude that the straight line path taken by our squeezing operator is the shortest path in the space of paths generated by the entire $SU(2) \times SU(2)$ algebra of generators. This is unsurprising: the Hamiltonian of the massless 3+1 theory in the chiral basis can be seen to be two copies of the 1+1 Dirac theory that we studied earlier, after the phase redefinitions etc. that we did. So the path length can be understood as path lengths in two separate factorised Hilbert spaces, each of which contains the $SU(2)$ structure.

\subsubsection{Massive Theory in 3+1 Dimensions}

Now we go on to consider the massive theory in 3+1 dimensions in the Dirac basis:
\bea
\gamma^{0}=\left(
\begin{array}{cc}
 I_{2} & 0 \\
 0 & -I_{2} \\
\end{array}
\right)
\quad
\gamma^{k}=\left(
\begin{array}{cc}
 0 & \sigma^{k} \\
 -\sigma^{k} & 0 \\
\end{array}
\right)
\eea
The Hamiltonian in terms of the Fourier variables is
\bea
H = \int d^{3}k  \overline{\Psi}(k)( k_{i}\gamma^{i} + m)\Psi(k) = \int d^{3}k \mathcal{H}_{k}
\eea
where $i$ runs from 1 to 3.
Interpreting the components of the spinor $\Psi(k)$ as oscillators, the Hamiltonian for the four oscillators at $k$ is
\bea
\mathcal{H}_{k} =  \frac{m}{2}[\Psi^{\dagger}_{1}, \Psi_{1}]+ \frac{m}{2}[\Psi^{\dagger}_{2}, \Psi_{2}]+ \frac{m}{2}[\Psi_{3}, \Psi^{\dagger}_{3}] +\frac{m}{2}[\Psi_{4}, \Psi^{\dagger}_{4}]+ \hspace{1.5in} \\ \nonumber + (k_{1}- i k_{2})  ( \Psi^{\dagger}_{1} \Psi_{4} + \Psi^{\dagger}_{3} \Psi_{2}) +  (k_{1}+i k_{2}) (\Psi^{\dagger}_{2} \Psi_{3} + \Psi^{\dagger}_{4} \Psi_{1})  + k_{3}( \Psi^{\dagger}_{1} \Psi_{3} -\Psi^{\dagger}_{2} \Psi_{4}+ \Psi^{\dagger}_{3} \Psi_{1} - \Psi^{\dagger}_{4} \Psi_{2}) 
\eea
At each $k$ let's define the operators
\bea 
b_{1} = e^{i \frac{\theta}{2}} \Psi_{1} \quad b_{2} = e^{-i \frac{\theta}{2}} \Psi_{2} \quad b_{3} = e^{-i \frac{\theta}{2}} \Psi^{\dagger}_{3}
 \quad b_{1} = e^{i \frac{\theta}{2}} \Psi^{\dagger}_{4}
\eea
where $\theta = \tan^{-1} {(\frac{k_{2}}{k_{1}})}$ and define $ \kappa = \sqrt{k^{2}_{1} + k^{2}_{2}}$. In terms of these the hamiltonian for the four oscillators at $k$ becomes
\bea
\mathcal{H}_{k} = \Big( \frac{m}{2} \sum^{4}_{i=1} [b^{\dagger}_{i}, b_{i}] \Big) 
+ k_{3} (b^{\dagger}_{1}b^{\dagger}_{3} - b_{1}b_{3}) + \nonumber k_{3} (b^{\dagger}_{4}b^{\dagger}_{2}- b_{4}b_{2}) + \kappa (b^{\dagger}_{2}b^{\dagger}_{3} - b_{2}b_{3}) + \kappa(b^{\dagger}_{1}b^{\dagger}_{4}- b_{1}b_{4})\\
\eea
After the usual B-V transformation this becomes
\bea
H = \int d^{3}k \frac{1}{2} \omega_{k} \sum^{4}_{i=1} [\tilde{b}^{\dagger}_{i}, \tilde{b}_{i}]
\eea
where $\omega_{k}=\sqrt{k^{2}_{1} + k^{2}_{2}+k^{2}_{3}+m^{2}}$.\\
The B-V transformation is
\bea
\left(
\begin{array}{c}
\tilde{b}_{1} \\
\tilde{b}_{2} \\
\tilde{b}_{3} \\
\tilde{b}_{4} \\
\tilde{b}^{\dagger}_{1} \\
\tilde{b}^{\dagger}_{2} \\
\tilde{b}^{\dagger}_{3} \\
\tilde{b}^{\dagger}_{4} \\
\end{array}
\right)=\frac{1}{ \sqrt {2 \omega_{k}({\omega_{k}-m})}}\left(
\begin{array}{c}
 b_{1} \kappa-b_{2} k_{3}+b^{\dagger}_{4} \left(\omega_{k}-m\right) \\
b_{2} \kappa+b_{1} k_{3}+b^{\dagger}_{3} \left(\omega_{k}-m\right) \\ 
-b_{3} \kappa+b_{4} k_{3}+b^{\dagger}_{2} \left(\omega_{k}-m\right) \\
-b_{4} \kappa-b_{3} k_{3}+b^{\dagger}_{1} \left( \omega_{k} - m\right)\\
b^{\dagger}_{1} \kappa-b^{\dagger}_{2} k_{3}+b_{4} \left(\omega_{k}-m\right)\\
b^{\dagger}_{2} \kappa+b^{\dagger}_{1} k_{3}+b_{3} \left(\omega_{k}-m\right)\\
-b^{\dagger}_{3} \kappa+b^{\dagger}_{4} k_{3}+b_{2} \left(\omega_{k}-m\right)\\
-b^{\dagger}_{4} \kappa-b^{\dagger}_{3} k_{3}+b_{1} \left(\omega_{k} -m \right)
\end{array}
\right)
\eea

We will take the reference state $|R\rangle$ is defined as the ground state of the ultra local Hamiltonian
\bea
H_{m} = \int d^{3}x \big(  m \overline{\Psi}(x) \Psi(x) \big)
=\int d^{3}k \big(  m \overline{\Psi}(k) \Psi(k) \big)=\int d^{3}k \frac{m}{2} \sum^{4}_{i=1} [b^{\dagger}_{i}(k), b_{i}(k)] \label{ulocal3d}
\eea
This state is annihilated by $b_{i}(k)$ for $i=$ 1,2,3,4 and $\forall k \in \R^{3}$. This state has no entanglement in $x$ space.
The target state is the approximate ground state $|T^{(\Lambda)}\rangle$ defined as
\bea
\tilde{b}_{i}(k)|T^{(\Lambda)}\rangle=0 \qquad \forall k : |k| \leq \Lambda \\ \nonumber
b_{i}(k)|T^{(\Lambda)}\rangle=0 \qquad \forall k : |k| > \Lambda
\eea

The basic observation of this section is that the target state can be reached from the reference state by the unitary transformation
\bea
|T^{(\Lambda)}\rangle=e^{-i \int_{|k| \leq \Lambda} d^{3}k r_{(k)} K(k)}| R \rangle=U|R\rangle
\eea
where the squeezing operator $K(k)$ is
\bea
\begin{split}
K(k) = i k_{3}(b^{\dagger}_{1}b^{\dagger}_{3} + b_{1}b_{3}) + i k_{3} (b^{\dagger}_{4}b^{\dagger}_{2}+ b_{4}b_{2}) + i \kappa (b^{\dagger}_{2}b^{\dagger}_{3} + b_{2}b_{3}) + i \kappa (b^{\dagger}_{1}b^{\dagger}_{4}+ b_{1}b_{4}) \label{squeezemassive3d}
\end{split}
\eea
and $r_{(k)}$ is
\bea
r_{(k)}=-\frac{1}{|k|} \arctan \Big( \frac{|k|}{m+ \omega_{k}} \Big). \label{massiverk}
\eea
Here $ |k|= \sqrt{k^{2}_{1} + k^{2}_{2}+k^{2}_{3}}$.

This unitary transformation takes the $b_{i}$ to a linear combination of $\tilde{b}_{i}$ via the similarity transformations
\bea
Ub_{1}U^{\dagger}=\frac{\kappa \tilde{b}_{1}+k_{3}\tilde{b}_{2}}{|k|} \qquad
Ub_{2}U^{\dagger}=\frac{k_{3}\tilde{b}_{1}-\kappa \tilde{b}_{2}}{|k|}\\ \nonumber
Ub_{3}U^{\dagger}=\frac{\kappa \tilde{b}_{3}+k_{3} \tilde{b}_{4}}{|k|}\ \qquad
Ub_{4}U^{\dagger}=\frac{k_{3} \tilde{b}_{3}-\kappa \tilde{b}_{4}}{|k|}
\eea
Crucially, there are no creation operators in the linear combinations on the right hand side, hence this unitary transformation takes the reference state to the target state. 

Now we can as usual consider an arbitrary path generated by the squeezing operator
 \bea
\begin{split}
|\Psi(\sigma)\rangle=e^{-i \int^{\sigma}_{0} ds \int_{|k| \leq \Lambda} d^{3}k  y_{k}(s) K(k)}| R \rangle 
=e^{-i \int_{|k| \leq \Lambda} d^{3}k Y_{k}(\sigma) K(k)}| R \rangle
\end{split}
\eea
Path ordering is not necessary because all $K(k)$ commute. Evaluating the length of this path using Fubini-Study metric gives 
\bea 
l(|\Psi(\sigma)\rangle)=\ \int_{0}^{1} d\sigma \sqrt{2 { \rm Vol} \int_{|k| \leq \Lambda} d^{3}k |k|^{2}
 (\partial_{\sigma} Y_{k} (\sigma))^{2}} 
\eea
The change of variable analogous to the massless case takes the form
\bea
X_{k}(\sigma)=|k|  Y_{k} (\sigma)
\eea
and the length of the path takes the usual form
\bea
l(|\Psi(\sigma)\rangle)=\ \int_{0}^{1} d\sigma \sqrt{2 { \rm Vol} \int_{|k| \leq \Lambda} d^{3}k 
 (\partial_{\sigma} X_{k} (\sigma))^{2}} 
\eea
This is again a flat Euclidean geometry associated with coordinate $X_{k} (\sigma)$ and so the geodesic is
\bea
X_{k}(\sigma) =  |k| r_{k}  \sigma
\eea
The complexity is the length of this path, and is given by
\bea
C = \sqrt{2 { \rm Vol} \int_{|k|\leq \Lambda} d^{3}k \Bigg( \arctan  \Big( \frac{|k|}{m+\omega_{k} } \Big) \Bigg)^{2} }
\eea
This integral can be explicitly evaluated:
\bea
\begin{split}
C^{2}  =\frac{1}{18} \pi  {\rm Vol} \left\lbrace12 i m^3 \text{Li}_2\left(1-\frac{2 m}{m-i \Lambda }\right) \right.\\\left. -24 m \left(\Lambda ^2+2 m^2 \log \left(\frac{2 m}{m-i \Lambda }\right)+m^2\right) \tan ^{-1}\left(\frac{\Lambda }{\sqrt{\Lambda ^2+m^2}+m}\right) \right.\\\left. +m^2 \left(12 \Lambda +i \pi ^2 m\right)+  48 \left(\Lambda ^3+i m^3\right) \left( \tan ^{-1}\left(\frac{\Lambda }{\sqrt{\Lambda ^2+m^2}+m}\right)\right)^2\right\rbrace
\end{split}
\eea
Its behaviour at large $\Lambda$ takes the form
\bea
C^{2}=\frac{\pi ^3}{6} {\rm Vol} \left(   \Lambda ^3 - \frac{6m}{\pi} \Lambda ^2  +\frac{12m^2}{\pi^{2}}  \Lambda  +\frac{4 m^3}{\pi} \log \left(\frac{\Lambda}{2m} \right)-\frac{2 m^3}{3\pi} + {\cal O}( 1/\Lambda)\right)
\eea 
When $m=0$, the exact form of the integral is simple to write down:
\bea
C^2 = \frac{\pi^3 \Lambda^3 {\rm Vol}}{6} \label{massless-massive}
\eea

Analogous to the previous cases, one can ask whether there exists a bigger class of generators in which our squeezing operator is a specific linear combination. It is straightforward to see that this is the case, and that the natural algebra generated by the creation annihilation operators is an $SO(8)$, and therefore a natural geometry that arises in the space of these more general paths is an $S^7$. We elaborate on this in an Appendix.

\section{Fermionic cMERA}

The discussion we have had in the previous section is very close in spirit to the so-called cMERA tensor network, which one can view in our language as just as an alternate choice of path connecting the reference state to the target state. The cMERA circuit can be viewed as entangling the neighbouring oscillators in $x$ space and doing a scale transformation iteratively. 

In what follows we will define the target state at finite cut-off to be some approximate ground state using cMERA. This approximate ground state need not be the same approximate ground state as defined previously. But as the cut-off is taken to infinity both will tend to the true ground state of the theory. Our discussion of cMERA will follow the papers \cite{cMERA, Takayanagi}. We will compute the length of the cMERA path for fermions. The discussions in \cite{cMERA, Takayanagi} are for 1+1 dimensional fermions, we will also consider a cMERA-like path that is a natural generalisation of these works to 3+1 dimensions. 

\subsection{Dirac cMERA in 1+1 Dimensions}

We will use the version of cMERA that is described for the Dirac theory in 1+1 dimensions in \cite{Takayanagi}. 

The cMERA path $|\Psi(u)\rangle$ is
\bea
|\Psi(u) \rangle = P e^{- i \int^{u}_{-\infty} du' \int
\ {dk K(k) y_{k}(u')}}| R \rangle  \label{cmerapsi}
\eea
where $|R\rangle$ is the reference state defined in equation \eqref{Dirac-Ref},  $K(k)$ is the squeezing operator given in equation \eqref{Dirac-squeeze} and \cite{Takayanagi}
\bea
y_{k}(u)=g(u) \frac{k  e^{-u}}{\Lambda} \Theta(\Lambda e^{u}-|k|)
\eea
where $\Theta(x)$ is the step function (it is $1$ for $x \geq 0$ and zero elsewhere) and $g(u)$ is
\bea
g(u) = \frac{1}{2} \Big( -\arcsin{ \frac{\Lambda e^{u}}{\sqrt{\Lambda^{2} e^{2u} + m^2}}} + \frac{m \Lambda e^{u}}{\Lambda^{2} e^{2u} + m^2} \Big).
\eea
Note that the integration range of $k$ in \eqref{cmerapsi} is superficially {\em not} restricted and is from $-\infty$ to $+\infty$. But effectively there are restrictions, arising from the step function.  It is important to be careful about this type  of thing in what follows, especially when changing the order of integration in $u$ and $k$ in the double integral. 
The path is parametrised from $-\infty$ to $0$ by the  parameter $u= \ln \sigma$, where $\sigma$ is our conventional path parameter that runs from $0$ to $1$. The parametric values of the reference state and target state is $-\infty$ and $0$ respectively. 

The target state reached by cMERA is
\bea
\begin{split}
|\Psi(0) \rangle = P e^{- i \int_{|k|\leq \Lambda}{dk K(k) Y_{k}(0)}}| R \rangle 
= P e^{- i \int_{|k|\leq \Lambda}{dk K(k) R_{k}}}| R \rangle  
\end{split}
\eea
where\footnote{Note that some of the expressions we present here are  superficially in tension with the results in \cite{Takayanagi}. This is because (we believe) the results --see in particular (121-123)-- in \cite{Takayanagi} should be compared only upto terms suppressed by $\Lambda$. A clean way to see this is to note that the $\phi_k$ in (121) and (122) in \cite{Takayanagi} do not have the same $\Lambda$ dependence: in particular, (122) vanishes when $|k|=\Lambda$, but (121) does not.} 
\bea
Y_{k}(0)=R_{k}(\Lambda)=\int^{0}_{-\infty} du y_{k}(u)=\Big( \frac{k}{2 \Lambda} \arcsin{\frac{\Lambda}{\sqrt{\Lambda^{2}+m^{2}}}} \Big) + r_{k}
\eea
where $r_{k}$ is given by eqn \eqref{rk1+1}. Note that changing the order of integration has brought out the promised $|k| \le \Lambda$ in the $k$-integral. This target state reproduces the true ground state in the $\Lambda \to \infty$ limit as $R_{k} \to r_{k}$ in this limit. But note that at finite values of the cut-off, it does {\em not} lead to the target state that we defined in our 1+1 dimensional discussion in the previous section. This is unlike in the case of the bosonic case that was considered in \cite{Chapman} where the cut-off target state had an extra parameter available (it was called $M$ in \cite{Chapman, Takayanagi}) and this could be used to make the two cut-off target states identical. We give a conceptual explanation for the absence of this extra scale in the fermionic case at the end of Appendix A, as a feature implicit in the structure of bosonic vs fermionic oscillators. The fact that the states are not the same at finite cut-off means that it is not meaningful to compare quantities at finite cut-off. We elaborate on some aspects of this observation in an Appendix.

At intermediate points on the cMERA path
\bea
|\Psi(u)\rangle=U(u)|R\rangle = e^{-i\int_{|k| \leq \Lambda e^u} dk K(k) Y_{k}(u)}|R\rangle \label{c-path}
\eea
with 
\bea 
Y_{k}(u)=\int^{u}_{-\infty} du' y_{k}(u')=\Big( \frac{k}{2 \Lambda e^{u}} \arcsin{\frac{\Lambda e^{u}}{\sqrt{\Lambda^{2} e^{2u}+m^{2}}}} \Big) + r_{k} \label{cMERAY}
\eea
Note again the $k$-integration range, again getting fixed by the change of integration order.
Evaluating the length of this cMERA path using Fubini-Study metric gives 
\bea
l_{cMERA}=\sqrt{\frac{2}{3}\Lambda {\rm Vol}}\int^{0}_{-\infty} du \ |(g(u)| \ e^{\frac{u}{2}} 
\eea
In some of the calculations, it is useful to note that \eqref{c-path} can be written as
\bea
|\Psi(u)\rangle=e^{-i\int_{|k| \leq \Lambda} dk K(k) Y_{k}(u)\Theta(e^u-|k|/\Lambda)}|R\rangle \equiv e^{-i\int_{|k| \leq \Lambda} dk K(k) Y^\theta_{k}(u)}|R\rangle
\eea
In any event, it is possible to calculate this length explicitly, and we present the result in the Appendix where we summarise various explicit formulas for complexities and circuit lengths. Here we merely note that when $m=0$ 
\bea
g(u)=-\frac{\pi}{4}
\eea
So the length of the cMERA path in this case is
\bea
l_{cMERA}=\sqrt{\frac{\pi^{2}\Lambda {\rm Vol}}{6}}
\eea

Now let us plot the $|\beta_{+}|$ for the following two paths. The straight line path taking the reference state to the cMERA target state is 
\bea
|\beta_{+}| = \tan{(R_{k} \sigma)} 
\eea
For the cMERA path, from \eqref{c-path}, \eqref{cMERAY} we find 
\bea
|\beta_{+}| = \tan\Big({\frac{ \arcsin{\frac{k}{\sqrt{k^2+m^2}}}}{2} - \frac{k \arcsin{\frac{\Lambda \sigma}{\sqrt{\Lambda^{2} \sigma^{2} +m^2}}}}{2\Lambda \sigma}}\Big)\Theta(\sigma-|k|/\Lambda)
\eea
where the step function arises because in calculating $\beta_+$ we want the entire $u$(or $\sigma$)-dependence to be on the integrand in \eqref{c-path}.
\begin{figure}
\centering
    \subfigure{\includegraphics[width=0.7\textwidth]{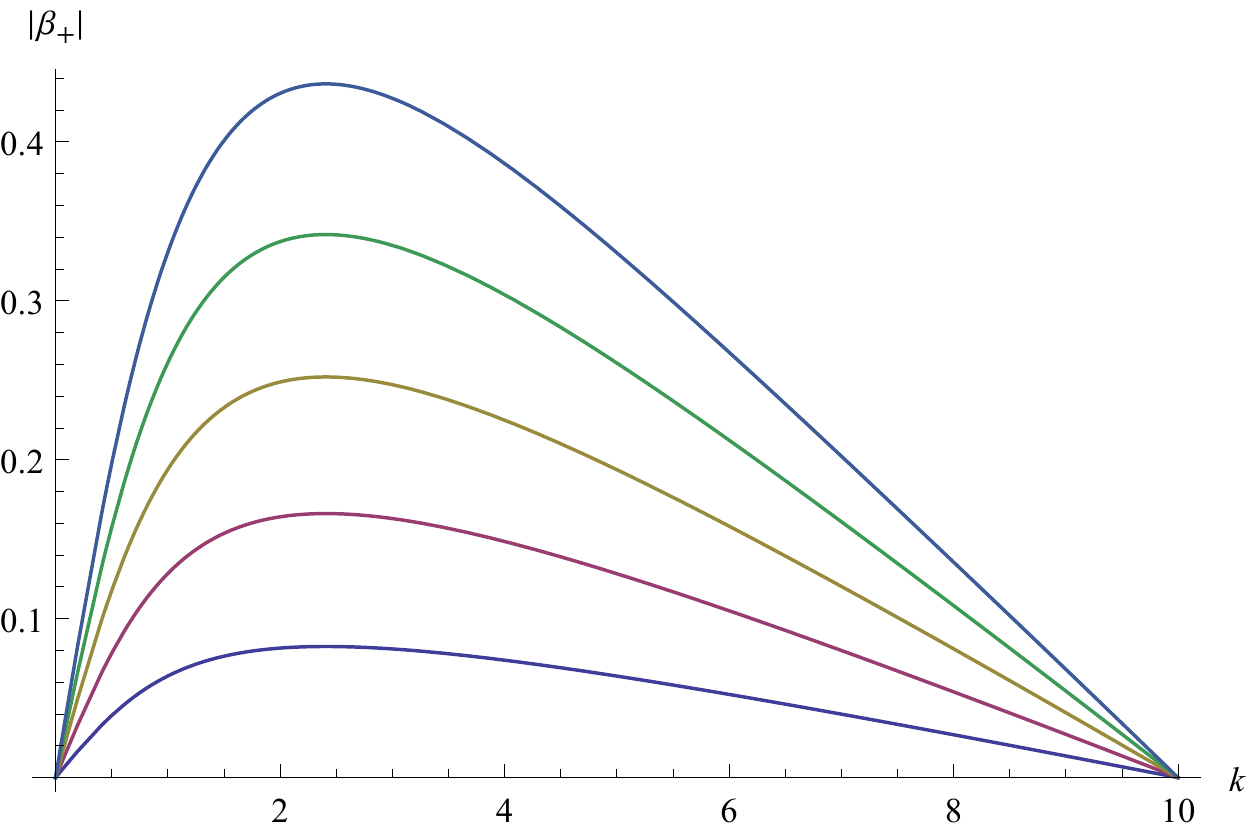}}
    
   \subfigure{\includegraphics[width=0.7\textwidth]{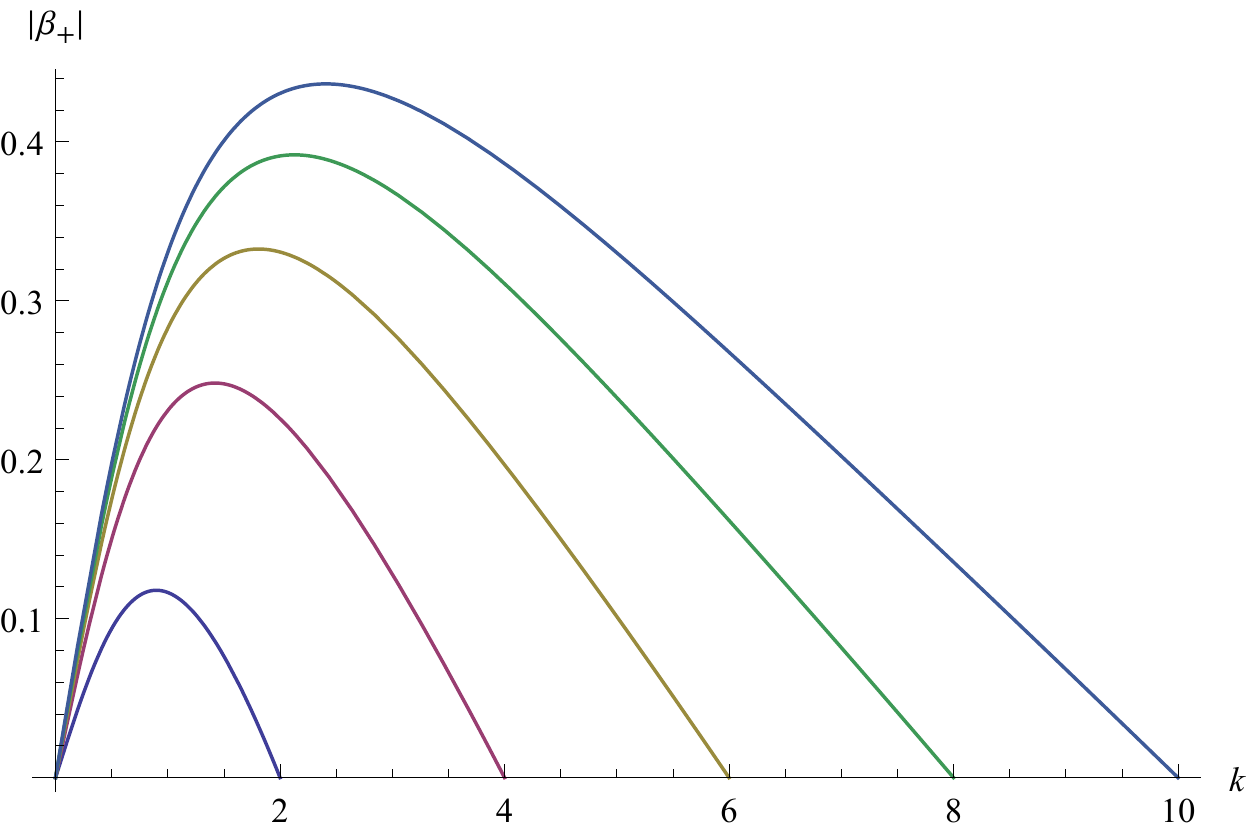}}
    
   \caption{The $|\beta_{+}|$ of first plot is for the geodesic path and the second plot is for the cMERA path. In both cases the target state is the cMERA target state. The above figures shows the plots of $|\beta_{+}|$ vs $k$ for $\Lambda=10$ and $m=1$ for the values of $\sigma=0.2,0.4,0.6,0.8,1.0$, increasing peak of the curve signifies increasing $\sigma$.}
    \label{plots}
\end{figure}

The $|\beta_{+}|$ of first (upper) plot here is for the geodesic path and the second plot is for the cMERA path. In both cases the target state is the cMERA target state.
The theta function is responsible for cutting off the curves in the second figure at the $k$-axis: in practice this means that the cMERA path $\beta_+$ has no support on $k > \Lambda \sigma$, whereas the straight line geodesic path has support on all $k$. This was observed in the bosonic case in \cite{Chapman} as well.

\subsection{3+1 Dimensions}

We can generalize an analogous construction to 3+1 dimensions (which we will call a cMERA-like path), and  calculate the length of a similar construction for fermions in 3+1 dimensions in the Dirac basis. The path $|\Psi(u)\rangle$ we take is
\bea
|\Psi(u) \rangle = P e^{- i \int^{u}_{-\infty} du' \int_{}{d^{3}k \ K(k) y_{k}(u')}}| R \rangle 
\eea
where $|R\rangle$ is the ground state of the ultra local Hamiltonian \eqref{ulocal3d},  $K(k)$ is the squeezing operator given in equation \eqref{squeezemassive3d} and 
\bea
y_{k}(u)=g(u) \frac{|k|  e^{-u}}{\Lambda} \Theta(\Lambda e^{u}-|k|)
\eea
and $g(u)$ is 
\bea
g(u) = -\frac{2}{\Lambda e^{u}} \arcsin{\frac{\Lambda e^{u}}{\sqrt{Z^{2}+\Lambda^{2} e^{2u}}}}
+\frac{ m \sqrt{Z}}{2(\Lambda^{2} e^{2u} +m^{2})^{\frac{3}{4}} \sqrt{mZ + \Lambda^{2} e^{2u}}}
\eea
with $Z \equiv \sqrt{\Lambda^{2} e^{2u}+m^{2}} +m $. To obtain $g(u)$ we follow the prescription of \cite{Takayanagi} and use
\bea
g(u)\equiv -\frac{|k|^2}{\Lambda}\frac{\partial}{\partial |k|}\Big(\frac{\Lambda\ r_k}{|k|}\Big)\Big|_{|k|=\Lambda e^u}
\eea
where $r_k$ is given by \eqref{massiverk}.

In the following, we will discuss the massless case for simplicity because that is enough to make our points. But the integrals in the above expressions are explicitly doable and we have calculated them also for the massive case. In particular, we have checked that the ground state is attained by the above path even in the massive case. 

In the $m=0$ case the $g(u)$ reduces to
\bea
g(u)= -\frac{\pi}{2 \Lambda e^{u}} \eea
The path is parametrised from $-\infty$ to $0$ by the  parameter $u$. The parametric values of the reference state and target state is $-\infty$ and $0$ respectively. The target state reached by this path is 
\bea
|\Psi(0) \rangle = P e^{- i \int_{|k|\leq \Lambda}{d^3 k K(k) Y_{k}(0)}}| R \rangle 
\eea
where $Y_{k}(0)$ is 
\bea
Y_{k}(0)=\int^{0}_{-\infty} du y_{k}(u) = \frac{\pi}{4} \big( \frac{k}{\Lambda^{2}} - \frac{1}{k} \big )
\eea
At finite $u$, we have
\bea
|\Psi(u) \rangle = P e^{- i \int_{|k|\leq \Lambda e^u}{d^3 k K(k) Y_{k}(u)}}| R \rangle = P e^{- i \int_{|k|\leq \Lambda}{d^3 k K(k) Y_{k}(u) \Theta(e^u-|k|/\Lambda)}}| R \rangle
\eea
where 
\bea
Y_k(u)=\frac{|k|}{\Lambda} \int_{\ln |k|/\Lambda}^u du \ e^{-u} g(u). 
\eea
Evaluating the length of this cMERA-like path using Fubini-Study metric gives 
\bea
l_{cMERA}=\sqrt{\frac{8}{7} \pi \Lambda^{5} {\rm Vol}}\int^{0}_{-\infty} du |g(u)| e^{\frac{5u}{2}} 
\eea
So the length of the cMERA-like path is
\bea
l_{cMERA}= \sqrt{\frac{8 \pi^{3} {\rm Vol} \Lambda^{3} }{63}}
\eea

One interesting feature of this construction is that one can see by comparing with \eqref{massless-massive} that at finite cut-off, the supposedly minimal complexity found in \eqref{massless-massive} is {\em higher} than the one found here. This is because the target states in both cases reach the true ground sate only at infinite cut-off, and it is not meaningful to compare the lengths to the two target states (which are distinct at finite cut-off). We present an example in 1+1 dimensions that clarifies and illustrates this type of cut-off dependence in an Appendix. 


\section{Future Directions and Speculations}

In this paper, we have calculated various natural notions of complexity in the space of unitary circuits for free fermionic quantum field theories. It seems possible that these results will be of some use in understanding the holographic significance of complexity (if any).  We will conclude in this section by listing various future directions beyond the ones we briefly touched upon in the introduction. 

One of the questions that might be of interest is to understand the physical content hiding behind a cut-off dependent quantity like complexity. We have discussed the question of ambiguities that arise due to the cut-off in the target states and in the path. It is an interesting questions what IR sensible physical quantities we can extract from these without knowing the full UV completion of these theories. There is some parallels here to the case of entanglement entropy. Entanglement entropy is also not an observable in the conventional sense like complexity, but it does capture interesting physical information that is sensible in the IR, even though it is UV-divergent \cite{ent-review}. It is tempting to also speculate whether supersymmetric theories with their better defined UV behaviour lead to any special simplifications when calculating these quantities. But note that such simplifications do not seem to happen for entanglement entropy.


Two obvious interesting directions from a holographic perspective is to introduce gauge invariances and interactions.  Perturbatively adding interactions seems reasonably straightforward, but ultimately for holographic purposes,  we would like to get to  a strongly coupled (conformal) gauge theory where we will probably need new ideas. The question of gauge invariance on the other hand seems like one where substantial progress might not be too hard to make: perhaps an approach along the lines of \cite{Soni} which uses Wilson line based variables to define entanglement entropy might be useful here as well for non-Abelian gauge theories. In the case of asymptotically free theories, it seems reasonable that the UV behaviour we find here for complexity will capture some of the relevant physics there as well. At least in an appropriate free theory limit, it seems plausible that the free field approach we have used might also for work for gauge theories once one takes care of ghost modes as well. A discussion of complexity in Abelian gauge theories was done in \cite{Abelian}. 

A simple situation which is quite interesting conceptually, and is a natural generalisation of the work we have done in this paper is to consider perturbative (bosonic and super) string theory on the worldsheet. Unlike in the case of entanglement entropy where target space issues complicate some of the questions \cite{Balasubramanian}, at least in the free limit, the circuit complexity of perturbative string theory should be conceptually easier to characterize and compute. This is because the reference states and target states can be understood purely within the worldsheet theory as long as one makes sure that the central charge is zero. Some work along this direction will be reported elsewhere. 

One thing we have ignored in this paper is the question of penalty factors for directions in circuit space \cite{Myers} and the possibility of more general Finsler/non-Riemannian metrics. At the moment, it is not clear to us what metrics are more natural than others, so we have restricted our attention to simple Riemannian choices. A related question that clearly needs a better understanding is the question of what qualifies as a natural choice of interesting gates when defining complexity. We have used some natural choices suggested by the problem itself in the discrete and continuum cases, but it will be nice to find a canonical way to choose these gates. 

We had two converging motivations for writing this paper. One was to understand complexity in quantum field theory as a pre-requisite for understanding holographic complexity \cite{coredump}. A second reason was the recent emergence of a type of strongly coupled fermionic theories called SYK (and related) models which have a controllable large-$N$ expansion. In particular, one possible case where a strongly coupled gauge theory could be solvable is in the context of the so-called SYK-like tensor models \cite{Witten}. These theories are not field theories, but they offer the possibility of exact solvability \cite{CK}, and therefore it might be possible to calculate complexity for them exactly. Between the insights one can get from free fermionic {\em field} theories we discussed here, and these strongly coupled gauged fermionic (and possibly holographic) quantum {\em mechanics}, it will be interesting to see if one can make any headway into an understanding of complexity in strongly coupled holographic gauge theories. 
 
\section*{Acknowledgments}

We thank Shadab Ahamed, Vishikh Athavale, Arpan Bhattacharyya, Aninda Sinha, Sudhir Vempati and Aditya Vijaykumar for discussions and/or correspondence. CK thanks the Dublin Institute for Advanced Studies (DIAS) for hospitality during the final stages of this work.

\appendix

\section{Oscillator Conventions}

We review some elementary facts to establish notation (as well as to set up a dictionary to go between notations). 

The phase space of the bosonic simple harmonic oscillator is comprised of two real variables $x$ and $p$. The number of degrees of freedom can be defined as half of the dimensionality of the phase space, and so the bosonic oscillator has exactly one degree of freedom. By combining the two phase space variables one defines the (complex, aka non-Hermitian) creation and annihilation operators $a$ and $a^\dagger$ 
\bea
a = \sqrt{\frac{\omega}{2}}x+i\frac{p}{\sqrt{2 \omega}}, \ \ 
a^\dagger =\sqrt{\frac{\omega}{2}}x - i\frac{p}{\sqrt{2 \omega}},
\eea
which satisfy the standard commutation relations 
\bea
[a, a^\dagger]=1, \label{ladder}
\eea
as a result of the canonical commutator, $[x, p]=i$. 
In terms of these the oscillator Hamiltonian can be written as 
\bea
H=\frac{1}{2}(p^2+\omega^2 x^2)=\omega\ (a^\dagger a +1/2)
\eea
where we have shifted away the zero point constant. We could have absorbed the $\sqrt{\omega}$'s in the definition of $a, a^\dagger$ above into the $x$ and $p$, without changing the discussion. But the factor of $\sqrt{2}$ in the definition can be absorbed only by introducing a compensating factor on (say) the right hand side of the canonical commutator: it changes the symplectic structure.

By analogy, the phase space of the simple fermionic oscillator will contain two real (but now Grassmann) variables $\psi_1$ and $\psi_2$. The number of degrees of freedom as given by half the phase space dimensionality is then 1. We define linear combinations of them which are again non-Hermitian
\bea
b= \frac{1}{\sqrt{2}}(\psi_1-i \psi_2), \ \ b^\dagger= \frac{1}{\sqrt{2}}(\psi_1+i \psi_2)
\eea
We want them to satisfy the anti-commutator 
\bea
\{ b, b^\dagger \} =1 \label{antiladder}
\eea
This is accomplished by the canonical anti-commutator (note that there is no $i$ on the right hand side):
\bea
\{\psi_i, \psi_j\}=\delta_{ij}, 
\eea 
The Hamiltonian can be taken again in analogy with the bosonic case
\bea
H =-i \omega \psi_1\psi_2 = \omega (b^\dagger b - 1/2)  
\eea
where the zero point constant now has (famously) the opposite sign.

A significant distinction between the fermionic and bosonic oscillators for our purposes is that for bosonic oscillators a scaling of the form
\bea
x \rightarrow \lambda x, \ p \rightarrow \lambda^{-1} p
\eea
for $\lambda \in \IC-\{0 \}$
preserves the commutation relations. Such a scaling is not possible in the fermionic case if one wants to preserve the anti-commutation relations. This operational fact is at the basis of our observation that we could not construct a reference state that depended on some arbitrary scale $M$ (like it was possible for bosons in \cite{Takayanagi, Chapman}) when dealing with fermions. It seems possible that this fact is of some deep significance, even outside the context of complexity, though we are not aware of a systematic exploration of it.   

\section{Bogoliubov-Valatin Transformations for Fermions}

Bogoliubov-Valatin (BV) transformations\footnote{We will only discuss the fermionic case here, an analogous discussion can be made for bosons as well, see \cite{BV}.} preserve the anti-commutation relations while diagonalising the Hamiltonian. Consider the quadratic Hamiltonian for some $n > 1$
\bea
H = \sum_{i,j =1}^{ n}{ \alpha_{i,j} [b^{\dagger}_{i},  b_{j}] + \frac{1}{2} \gamma_{ij} b^{\dagger}_{i} b^{\dagger}_{j} + \frac{1}{2} \gamma^{*}_{ij} b_{i}b_{j}} \label{generalH}
\eea
Where $ b_{i} $ and $ b^{\dagger}_{j} $ are the annihilation and creation operators respectively and follow the anti-commutation relations. To ensure that the Hamiltonian is Hermitian, the $ \alpha_{i,j} $ and $ \gamma_{i,j} $ must satisfy
\bea
\alpha_{ij} = \alpha^{*}_{ji} \quad \gamma_{ij} = - \gamma_{ji}
\eea 
Now, \eqref{generalH} can be written as
\bea
H = \frac{1}{2} \Psi^{\dagger}M \Psi
\eea
Where $ \Psi $ and $ \Psi^{\dagger} $ are 
\bea
\Psi = 
\left [\begin{array}{c}
b \\
(b^{\dagger})^T
\end{array}
\right ] 
\qquad 
\Psi^{\dagger} =
\left [\begin{array}{cc}
b^{\dagger} & b^T
 \end{array}
\right]
\eea
And $ b , (b^{\dagger})^T$ are column vectors of size $n$:
\bea
b = 
\left [ \begin{array} {c}
b_{1} \\
 b_{2} \\
 \vdots 
\\b_{n} 
\end{array} 
\right]
 \qquad 
(b^{\dagger})^T = 
\left [ \begin{array}{c}
b^{\dagger}_{1} \\
 b^{\dagger}_{2} \\
 \vdots \\
b^{\dagger}_{n} 
\end{array} \right ] \label{dagger-def}
\eea
The matrix $M$ has the form
\bea
M= 
\left [ \begin{array} {cc}
\alpha & \gamma \\ \gamma^{\dagger} & - \alpha^{T}
\end{array} \right ]
\eea
Our goal is to diagonalise the Hamiltonian and identify the normal modes, while preserving the anti-commutators. This is accomplished by the B-V transformations $T$:
\bea
\Psi = T \Phi,
\eea
Here, $ \Phi $ is the column matrix of normal modes:
\bea
\Phi = \left [ \begin{array} {c}
\tilde{b} \\ (\tilde{b^{\dagger}})^T \end{array} \right]
\eea
with
\bea
\tilde{b} = 
\left [ \begin{array} {c}
\tilde{b_{1}} \\
 \tilde{b_{2}} \\
 \vdots 
\\ \tilde{b_{n}} 
\end{array} 
\right]
 \qquad 
(\tilde{b^{\dagger}})^T = 
\left [ \begin{array}{c}
\tilde{b^{\dagger}_{1}} \\
 \tilde{b^{\dagger}_{2}} \\
 \vdots \\
\tilde{b^{\dagger}_{n}} 
\end{array} \right ]
\eea
To do the diagonalisation, $T$ must satisfy 
\bea
 D = T^{\dagger} M T \label{diagonal}
\eea
where $D$ is the diagonal matrix of normal mode frequencies. This is immediate from
\bea
H = \Psi^{\dagger} M \Psi  = \Phi^{\dagger} T^{\dagger} MT \Phi  = \Phi^{\dagger} D \Phi.
\eea
A general linear transformation is of the form
\bea
b_{i} = A_{ij} \tilde{b_{j}} + B_{ij} \tilde{b^{\dagger}_{j}}
\eea
which can be written in the matrix form 
\bea
 \left [\begin{array}{c}
b \\
(b^{\dagger})^T
 \end{array}
\right ] = 
\left [ \begin{array}{cc} A & B \\ B^{*} & A^{*}  \end{array} \right ]
\left [\begin{array}{c}
\tilde{b} \\
(\tilde{b^{\dagger}})^T
 \end{array}
\right ] \label{bogo}
\eea
This means that $T$ looks like
\bea
T = \left [ \begin{array}{cc} A & B \\ B^{*} & A^{*}  \end{array} \right ]\label{T}
\eea
Using the above explicit form of $T$, a direct calculation shows that the condition for $T$ to unitary is identical to the condition that the anti-commutators are preserved under \eqref{bogo}. It is useful when doing this calculation to remember manipulations like the following: for column vectors $x$ and $y$ if $x=A y$, ie., $x_i=A_{ij} y_j$, then 
\bea
x^\dagger_j=(A y)^\dagger_j=(y^\dagger A^\dagger)_j = y^{\dagger}_i A^{\dagger}_{ij}= y^\dagger_i A^*_{ji} = A^*_{ji} y^\dagger_i,
\eea 
and therefore, $(x^\dagger)^T=A^* (y^\dagger)^T$. Note that the transposition in this last equation is necessary because we want  to interpret it as a column vector equation, as in the second definition in \eqref{dagger-def}.

In conclusion, a B-V transformation for fermions is a unitary matrix $T$ of the form \eqref{T}, that can diagonalise the Hamiltonian (as captured by \eqref{diagonal}). The crucial point is that any unitary matrix that does the job of diagonalisation does not qualify as a B-V matrix\footnote{Note that the Hamiltonian being Hermitian, it can always be diagonalised by a unitary.}, it has to be of the form \eqref{T}.

\section{Bogoliubov-Valatin vs Dirac Modes}

The spinor $\Psi(x)$ was Fourier transformed in section 3.1 as
\bea
\Psi(x)=\int \frac{dk}{\sqrt{2 \pi}} \Psi(k) e^{i k x}. \label{C1}
\eea
A more standard basis is to expand in the basis of Dirac modes,
\bea
\Psi(x)=\int \frac{dk}{\sqrt{2 \pi}}  \frac{1}{\sqrt{2 E}} (a_{k} u_{k} +b^{\dagger}_{-k} v_{-k})e^{i kx} \label{C2}
\eea
where $a_{k}$, $b_{k}$ are the annihilation operators of the ground state and $u_{k}$ and $v_{k}$ are the spinors satisfying the equation
\bea
(\slashed{k}+m)u_{k}=0 \quad {\rm and} \quad (\slashed{k}-m)v_{k}=0
\eea
with the normalisation 
\bea
\bar{u}u=2m \quad {\rm and} \quad \bar{v}v=-2m.
\eea
These can be explicitly calculated to be
\bea
u_{k}= 
\left (\begin{array}{c}
\sqrt{E+m} \\
\sqrt{E-m} 
\end{array}
\right )
\quad {\rm and} \quad
v_{-k}= 
\left (\begin{array}{c}
-\sqrt{E-m} \\
\sqrt{E+m} 
\end{array}
\right )
\eea
Comparing \eqref{C1} and \eqref{C2} we get
\bea
\left (\begin{array}{c}
\Psi_{1}(k) \\
\Psi_{2}(k) 
\end{array}
\right )=\frac{1}{\sqrt{2 E}} (a_{k} u_{k} +b^{\dagger}_{-k} v_{-k})
\eea
from which
\bea
\begin{split}
a_{k}&=&A_{k} \Psi_{1} (k) + B_{k} \Psi_{2} (k) \\
b^{\dagger}_{-k}&=&-B_{k} \Psi_{1} (k) + A_{k} \Psi_{2} (k) 
\end{split}
\eea
follows, with the $A_{k}$ and $B_{k}$ given in section 3.1. It is clear from this that the ground state obtained there via a Bogoliubov-Valatin transformation in the main body of the paper, is just the usual ground state (as it should be).

We have phrased this discussion in 1+1 dimensions, but an entirely similar discussion holds in higher dimensions as well. The only difference is that in higher dimensions (say 3+1 d) the Dirac $u, v$ modes are not uniquely fixed (ie., the $u$ and $v$ come with an index as is familiar in 3+1 d where the index has two values), and so there are some more arbitrary choices that need to be made in the choice of their specific form. 

\section{Geodesics on $\IC\IP^1$}

The metric we derived in section 3.3 is the Fubini-Study metric on $\IC \IP^1$: 
\bea
ds^{2} = \frac{\beta'^{*}_{+} \beta_{+}'}{(1 + |\beta_{+}|^{2})^{2}}d \sigma^{2}
\eea
where $\beta'_{+}$ is $ \frac{d \beta_{+}}{d \sigma} $. Since $\beta_{+}$ is a complex variable, we can write $\beta_{+} = x + iy $, to bring the metric to a real form\footnote{Said another way, manifesting the complex structure of the metric is not necessary to answer metric questions.}:
\bea
ds^{2} = \frac{dx^{2}+ dy^{2}}{(1 + x^{2} + y^{2})^{2}}
\eea
The geodesics on this metric satisfy the usual Euler-Lagrange equations
\bea
\frac{d^{2} x^{\alpha}}{d \sigma^{2}} + \Gamma^{\alpha}_{\beta \gamma} \frac{d x^{\beta}}{d \sigma} 
\frac{d x^{\gamma}}{d \sigma} = 0
\eea
where $x^\alpha\equiv (x(\sigma),y(\sigma))$. For the straight line path given in \eqref{straightbeta}, $\beta_+$ takes the form
\bea
\beta_{+} = -i \tan{r_{k} \sigma}
\eea
So we get $x =0$ , $y = -\tan{r_{k} \sigma}$. This can immediately checked to satisfy the above geodesic equation.

Now for alternate path generated by $B(k)$ , the form of $\beta_+
$ is given in \eqref{B-beta}:
\bea
\beta_{+} = \frac{i \sin{r_{k} \sin{\frac{\pi \sigma}{2}}}}{i \cos{\frac{\pi \sigma}{2}} - \cos{r_{k}} \sin{\frac{\pi \sigma}{2}}}
\eea
This gives
\bea
x = \frac{\sin{r}\sin{\frac{\pi \sigma}{2}} \cos{\frac{\pi \sigma}{2}}}
{\cos^{2}{r} \sin^{2}{\frac{\pi \sigma}{2}} + \cos^{2}{\frac{\pi \sigma}{2}}}  \qquad y = \frac{\sin{r}\cos{r} \sin^{2}{\frac{\pi \sigma}{2}}}{\cos^{2}{r} \sin^{2}{\frac{\pi \sigma}{2}} + \cos^{2}{\frac{\pi \sigma}{2}}}
\eea
Plugging this into the left hand side of the geodesic equation above, we find that the equation is {\em not} satisfied.  

We see that straight line path generated by squeezing operator $K(k)$ is the geodesic of the $\IC \IP^1$ metric but the path taken by $B(k)$ is not.

\section{$SO(8)$}

In this appendix, we consider the massive 3+1 dimensional fermions discussed in the main text and consider a natural, yet more general class of generators that can be used as the primitive gates. The squeezing operator we constructed in the main body of the paper is a specific linear combination of these generators and can be viewed as a restricted class of paths. 

The basic observation is that the most general (quadratic) generators that we can construct at each $k$ from the creation and annihilation operators are built from
\bea
b^{\dagger}_{i} b^{\dagger}_{j}, \quad b_{i} b_{j}, \quad b^{\dagger}_{i} b_{j}
\eea
For $ i,j =1,2,3,4 $. Using the antisymmetry property,
 $ b^{\dagger}_{i}b^{\dagger}_{j}= -b^{\dagger}_{j}b^{\dagger}_{i}  $ and $ b_{i}b_{j}= -b_{j}b_{i}  $for $i \neq j$ and $ b_{i}b_{i}= b^{\dagger}_{i}b^{\dagger}_{i} =0 $, we can count that there are 6 + 6 +16 = 28 total generators. This is a strong suggestion that the algebra of these generators forms an $SO(8)$ algebra, which also has the same number of generators. 
 
To explicitly see the algebra, all one has to do is  define
\bea
\Gamma_{2i-1} = b_{i} + b^{\dagger}_{i}  \qquad 
\Gamma_{2i} = i(b_{i} - b^{\dagger}_{i})
\eea
and note that it follows from the anti-commutation relations that the $ \Gamma$'s satisfy the Clifford algebra
\bea
\{ \Gamma_{a}, \Gamma_{b} \} = 2 \delta_{ab}
\eea
for $SO(8)$.  The generators of $SO(8)$ are then as usual defined via
\bea
\Sigma_{ab} = -\frac{i}{4} [\Gamma_a, \Gamma_b]
\eea
Where $ a, b$ runs from 1 to 8. $\Sigma_{ab} $ is an antisymmetric tensor, which again yields the 28 independent generators of $SO(8)$. The fact that the $\Gamma_a$ defined from the $b_i$ and  $b_i^{\dagger}$ satisfy the Clifford algebra guarantees that the generators corresponding to $\Gamma_{(a} \Gamma_{b)}$ are proportional to the identity (when $a=b$, else they are zero) and therefore only contribute to a decoupled phase. 

In analogy with the 1+1 dimensional case where we found an $SU(2)$ algebra and an associated $\IC \IP^1$ geometry on which it naturally acts, this indicates that the natural geometry arising from the paths here is an $S^7$ where the $SO(8)$ acts linearly. It will be interesting to see whether the squeezing operator we have defined in the text leads to a geodesic or non-geodesic path when considered as a path in this geometry. Based on genericity alone, there is no reason why the squeezing operator path is geodesic, but it will be nice to (dis)prove this explicitly. The calculation will be the generalisation of the one we did for the $SU(2)$ case in 1+1 dimensions, but unlike the 3 generators there, now we have 28 generators. So one has to either go for more brute force than we have been able to muster, or find an alternate clever way to do this, or a suitable combination of both.

\section{Cut-off Dependence}

Notice that in section 4.2 the $\Lambda$-dependent length of the cMERA-like path is shorter than that of the straight line path calculated in section 3. This is a result of the fact that at finite cut off, the target states do not match and therefore it is not meaningful to compare them, even though they do match at infinite cut-off.


Let us define a one parameter family of target states for the 1+1 dimensional Dirac theory, with a cutoff $\Lambda$ such that the target state approaches the ground state of the field theory in the $\Lambda \rightarrow \infty$ limit:
\bea
|T^{(\Lambda)}_{\alpha}\rangle=e^{-i \int_{|k|\leq \Lambda} K(k) r(k) \big(1+\alpha \frac{k^{2}}{\Lambda^{2}} \big)} |R\rangle
\eea
where $r(k)$ is given in equation \eqref{rk1+1} and $\alpha \in \R$. In the limit $\Lambda \to \infty$ the coefficient of $K(k)$ in the exponent becomes $r(k)$ thereby reproducing the ground state of the field theory.

We will use the same reference state as the one we used in the main body of the paper and calculate the length of the minimal path from this reference state to this new target state. To simplify the integrals we will do this in the $m=0$ case. Again using Fubini-Study metric and following the by now standard derivation, we get the straight line path to be the minimal path. The length of this minimal path is
\bea
l_{\alpha}=\sqrt{{\rm Vol} \int_{|k|\leq \Lambda} r(k)^2 \left(1+\alpha \frac{k^{2}}{\Lambda^{2}} \right)^2}
\eea
For $m=0$
\bea
\begin{split}
l_{\alpha}=\sqrt{\frac{\pi^{2}}{8}\Lambda {\rm Vol} \Big( 1+\frac{2}{3}\alpha +\frac{1}{5}\alpha^{2}\Big)}\\
= C\Big( 1+\frac{2}{3}\alpha +\frac{1}{5}\alpha^{2}\Big)^{\frac{1}{2}}
\end{split}
\eea
where $C=\sqrt{\frac{\pi^{2}}{8}\Lambda {\rm Vol}}$ is the complexity that we calculated in the text.

It is easy to see that this length can be bigger or even smaller (eg: $\alpha =-5/3$) than the complexity $C$, depending on the value of $\alpha$. What is happening here is that the cut-off dependence in the definition of the target state is making the comparison between path lengths at finite cut-off meaningless between states that are distinct at finite cut off. This is what is manifest in section 4.2. There also the target states were different and so we do not expect the cMERA-like lengths to be directly comparable to the complexity at finite cut off. In section 4.1 it turned out that cMERA length was longer than the complexity and in section 4.2 it is the opposite. 

This is a demonstration of the fact that UV sensitive quantities are not stable under change of cut-off. It is interesting to ask what kind of information one can extract from a UV sensitive quantity like complexity: note that entanglement entropy is also typically UV divergent, but its sub-leading behaviour is proportional to the area. So it will be interesting to understand what kind of UV-insensitive information one can extract from quantities like complexity.

\section{Complexities and Circuit Lengths: Selected Summary}

\subsection{Dirac Complexity in 1+1 Dimensions}
The following is a real quantity
\bea
\begin{split}
C^{2}=-\frac{m}{2} {\rm Vol} \log \left(\frac{\Lambda ^2}{m^2}+1\right) \tan ^{-1}\left(\frac{\Lambda }{m}\right)-\frac{1}{4} m {\rm Vol} \left(-\log (16) \tan ^{-1}\left(\frac{\Lambda }{m}\right)+ \right.\\\left. -i \text{Li}_2\left(-e^{-2 i \tan ^{-1}\left(\frac{\Lambda }{m}\right)}\right)+i \text{Li}_2\left(-e^{2 i \tan ^{-1}\left(\frac{\Lambda }{m}\right)}\right)\right)+\frac{1}{2} \Lambda  {\rm Vol} \left(\tan ^{-1}\left(\frac{\Lambda }{m}\right)\right)^2
\end{split}
\eea
which gives
\bea
C^{2}=\frac{\pi ^2}{8}  {\rm Vol}\left(\Lambda +\frac{4 m}{\pi}\log \left(\frac{2 m}{\Lambda }\right)-\frac{4 m}{\pi}+ {\cal O}( 1/\Lambda)\right)
\eea

\subsection{Circuit Length with the $B$ Generator in 1+1 Dimensions}

\bea
C^{2}=\left(\frac{\pi }{2}\right)^2 {\rm Vol} \left(\Lambda +\frac{m}{2}  \log \left(\frac{\sqrt{\Lambda ^2+m^2}-\Lambda }{ \sqrt{\Lambda ^2+m^2}+\Lambda}\right)\right)
\eea
which gives
\bea
C^{2}=\frac{\pi ^2}{4}   {\rm Vol} \left( \Lambda + m\log \left(\frac{m}{2\Lambda}\right) + {\cal O}( 1/\Lambda)  \right)
\eea

\subsection{Massive Dirac Complexity in 3+1 Dimensions }
The following is again a real quantity,
\bea
\begin{split}
C^{2}  =\frac{1}{18} \pi  {\rm Vol} \left\lbrace12 i m^3 \text{Li}_2\left(1-\frac{2 m}{m-i \Lambda }\right) \right.\\\left. -24 m \left(\Lambda ^2+2 m^2 \log \left(\frac{2 m}{m-i \Lambda }\right)+m^2\right) \tan ^{-1}\left(\frac{\Lambda }{\sqrt{\Lambda ^2+m^2}+m}\right) \right.\\\left. +m^2 \left(12 \Lambda +i \pi ^2 m\right)+  48 \left(\Lambda ^3+i m^3\right) \left( \tan ^{-1}\left(\frac{\Lambda }{\sqrt{\Lambda ^2+m^2}+m}\right)\right)^2\right\rbrace
\end{split}
\eea
and it yields
\bea
C^{2}=\frac{\pi ^3}{6} {\rm Vol} \left(   \Lambda ^3 - \frac{6m}{\pi} \Lambda ^2  +\frac{12m^2}{\pi^{2}}  \Lambda  +\frac{4 m^3}{\pi} \log \left(\frac{\Lambda}{2m} \right)-\frac{2 m^3}{3\pi} + {\cal O}( 1/\Lambda)\right)
\eea

\subsection{cMERA Circuit Length in 1+1 Dimensions}

For the massive case, the explicit expression for the cMERA circuit length is
\bea
l_{cMERA}=\sqrt{\frac{{\rm Vol}}{6}} \left(2 \sqrt{\Lambda } \cot ^{-1}\left(\frac{m}{\Lambda }\right)+3  \sqrt{i m} \left(\tanh ^{-1}\left(\sqrt{\frac{i \Lambda}{m}}\right)-\tan ^{-1}\left( \sqrt{\frac{i \Lambda}{m}}\right)\right)\right) \eea
and it has the large $\Lambda$ form
\bea
l^{2}_{cMERA}=\frac{\pi^2}{6}   {\rm Vol} \left(   \Lambda -3    \sqrt{2m \Lambda }+\frac{9 m}{2}+\frac{8 m}{\pi}+ {\cal O}( 1/\Lambda) \right)
\eea

\end{document}